\documentclass[twocolumn]{aastex62}

\def\lapp{\ifmmode\stackrel{<}{_{\sim}}\else$\stackrel{<}{_{\sim}}$\fi}
\def\gapp{\ifmmode\stackrel{>}{_{\sim}}\else$\stackrel{>}{_{\sim}}$\fi}
\usepackage{graphicx}
\usepackage{multirow}
\usepackage{color}
\usepackage{amsmath}
\usepackage{soul}
\usepackage{hyperref}

\def\amin{\ifmmode^{\prime}\else$^{\prime}$\fi}
\def\asec{\ifmmode^{\prime\prime}\else$^{\prime\prime}$\fi}
\def\simgt{\lower.5ex\hbox{$\; \buildrel > \over \sim \;$}}
\def\simlt{\lower.5ex\hbox{$\; \buildrel < \over \sim \;$}}

\shorttitle{}
\shortauthors{}
\begin{document}

\title{A broadband X-ray study of the Rabbit pulsar wind nebula powered by PSR~J1418$-$6058}

\correspondingauthor{Hongjun An}
\email{hjan@cbnu.ac.kr}

\author{Jaegeun Park}
\author{Chanho Kim}
\affiliation{Department of Astronomy and Space Science, Chungbuk National University, Cheongju, 28644, Republic of Korea}
\author{Jooyun Woo}
\affiliation{Columbia Astrophysics Laboratory, Columbia University, New York, NY 10027, USA}
\author{Hongjun An}
\affiliation{Department of Astronomy and Space Science, Chungbuk National University, Cheongju, 28644, Republic of Korea}
\author{Kaya Mori}
\affiliation{Columbia Astrophysics Laboratory, Columbia University, New York, NY 10027, USA}
\author{Stephen P. Reynolds}
\affiliation{Physics Department, NC State University, Raleigh, NC 27695, USA}
\author{Samar Safi-Harb}
\affiliation{Department of Physics and Astronomy, University of Manitoba, Winnipeg, MB R3T 2N2, Canada}

\begin{abstract}
        We report on broadband X-ray properties of the Rabbit pulsar wind nebula (PWN)
associated with the pulsar PSR~J1418$-$6058 
using archival Chandra and XMM-Newton data, and a new NuSTAR observation.
NuSTAR data above 10 keV allowed us to detect the 110-ms spin period
of the pulsar, characterize its hard X-ray pulse profile,
and
resolve hard X-ray emission from the PWN after removing contamination
from the pulsar and other overlapping point sources. The extended PWN was detected up to
$\sim$20\,keV and is well described by a power-law model with a photon index
$\Gamma\approx$2.
The PWN shape does not vary significantly with energy,
and its X-ray
spectrum shows no clear evidence of softening away from the pulsar. 
We modeled the spatial profile of X-ray spectra and broadband spectral energy distribution
in the radio to TeV band to infer the physical properties of the PWN.
We found that a model with low magnetic field strength ($B\sim 10$ $\mu$G)
and efficient diffusion ($D\sim 10^{27}$ cm$^2$\,s$^{-1}$) fits the PWN data well.
The extended hard X-ray and TeV emission, associated respectively with synchrotron radiation
and inverse Compton scattering by relativistic electrons, suggests that particles
are accelerated to very high energies ($\gtrsim500$\,TeV),
indicating that the Rabbit PWN is a Galactic PeVatron candidate. 
\end{abstract}

\bigskip
\section{Introduction}
\label{sec:intro}
Pulsar wind nebulae (PWNe) are 
bubbles of relativistic particles powered
by the rotational energy released from a pulsar and 
generally emit electromagnetic radiation
from radio to TeV gamma-ray energies \citep[see][for a review]{Slane2017}.
It is widely accepted that the pulsar wind particles are accelerated to
very high energies at a termination shock \citep[][]{kc84a},  
and their interaction with magnetic fields and
ambient low-energy photons result in broadband emission from the radio to gamma-ray band.
In particular, 
the detection of many PWNe in the very high-energy
(VHE; $>$0.1TeV) gamma-ray band \citep[e.g.,][]{HESS2018} 
suggests that 
these sources are strong Galactic PeVatron candidates --  the particle acceleration and transport mechanisms in PWNe are essential for understanding the origin of TeV--PeV cosmic-ray electrons and positrons detected on Earth \citep[e.g.,][]{Fiori2022}. 
Furthermore, multi-wavelength studies of PWNe can provide insights into relativistic
shock physics \citep[e.g.,][]{skl15} and magnetohydrodynamic (MHD) flow
of high-energy particles \citep[e.g.,][]{kc84a}.

\begin{figure*}
\centering
\hspace{0 mm}
\includegraphics[width=7 in]{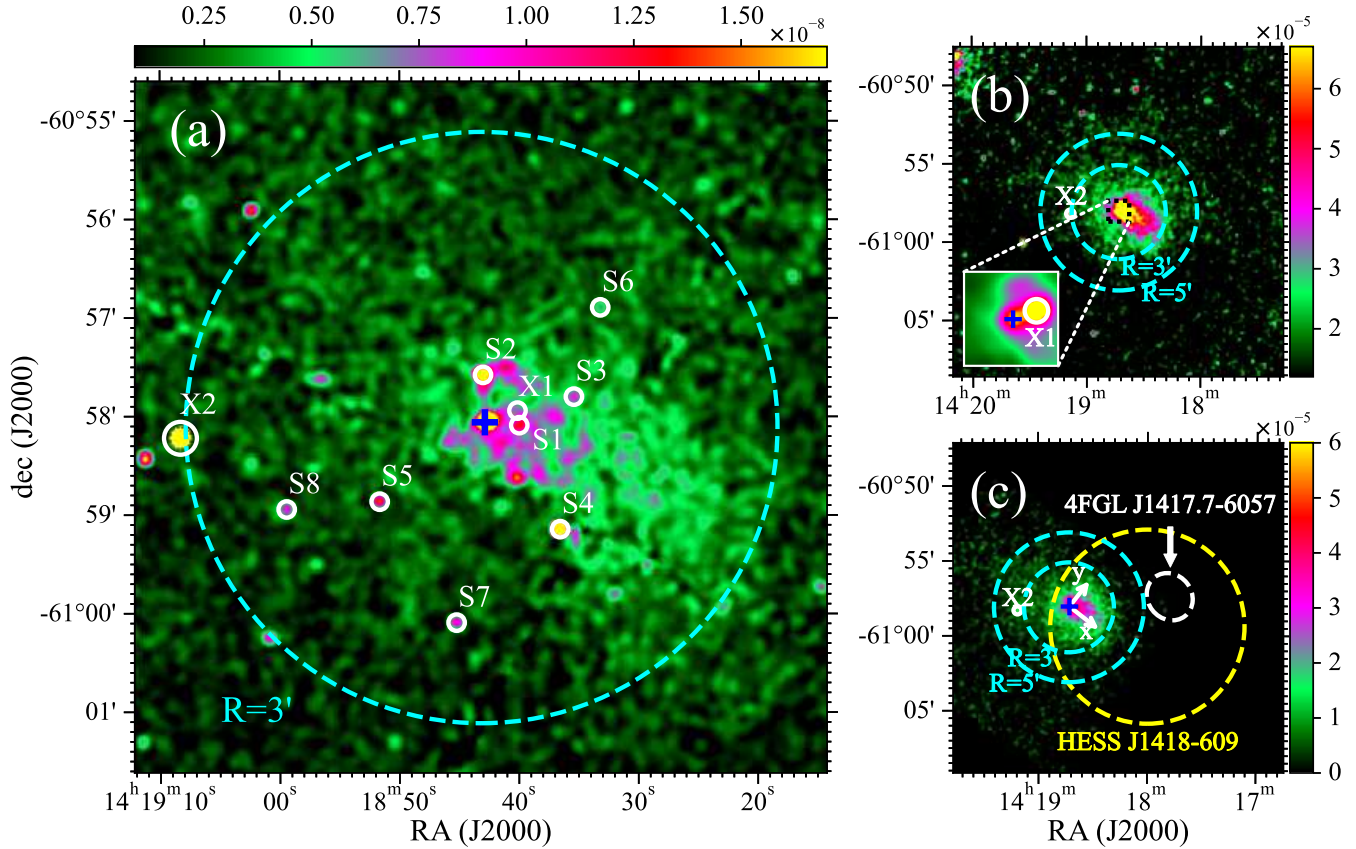} 
\figcaption{Background-subtracted and exposure-corrected X-ray images of the Rabbit PWN
measured with Chandra (a), XMM-Newton (b), and NuSTAR (c). The inset in panel (b) is a zoomed-in image of the central region.
The pulsar (J1418) position is marked by a blue cross, and 
contaminating point sources ($R\lapp 3'$) are denoted in white circles. The GeV and TeV counterparts of the PWN are
denoted in a white and a yellow ellipse in panel (c), respectively, and $R=3'$ and $R=5'$ circles (cyan dashed) are shown for reference.
Notice that X1 (9$''$ north of S1; panels (a) and (b)) was detected in the XMM-Newton image but not in the Chandra image.
The figures are smoothed, and scales are adjusted for better legibility.
\label{fig:fig1}}
\vspace{0mm}
\end{figure*}

	Broadband spectral energy distributions (SEDs) of PWNe are well characterized
by two components originating from the same population of relativistic particles:
synchrotron radiation ($E_\gamma\le$ MeV) and inverse Compton scattering (ICS)
of ambient photon fields in the gamma-ray band \citep[e.g.,][]{bb2003}.
Because the energy distribution of particles is imprinted in
the emission spectra, both obtaining and modeling SED data accurately allow us to
understand the particle acceleration processes in PWNe. 
Characterizing the synchrotron emission through imaging and spectroscopic data
in the hard X-ray band ($E_\gamma > 10$ keV) is particularly useful since it allows us to directly probe the highest energy (sub-PeV) particle distributions,  whereas the ICS component 
is mainly affected by the properties of the ambient photon fields.

In general, young rotation-powered PWNe ($\lesssim$$10^3$ yr) such as the Crab nebula and G21.5$-$0.9,
compared to older nebulae, are bright in X-rays and relatively faint in the VHE band, implying that magnetic fields
in young PWNe are likely strong \citep[$B\sim100\mu\rm G$;][]{Meyer2010,Guest2019}.
Particles in these PWNe lose energy efficiently via
synchrotron radiation, which is observed as a spectral break and/or a PWN size decrease
with increasing photon energy \citep[e.g.,][]{Reynolds2016}. The X-ray-to-VHE flux ratios of numerous PWNe
in different evolutionary stages (e.g., young, middle-aged and relic PWNe) have been  observed to decrease with their ages \citep[][]{Kargaltsev2013}. It implies that magnetic fields are lower 
\citep[e.g., $B\le$ 10$\mu\rm G$;][]{Kargaltsev2013,Zhu2018} in the older PWNe ($> 10^4$ yr) that can supply
the interstellar medium (ISM) with highly energetic particles
due to the weaker synchrotron cooling rates   \citep[e.g.,][]{Giacinti2020}.
	
	The Rabbit PWN was discovered in an X-ray study of a region around a bright EGRET source \citep[][]{Roberts1998}.
A radio counterpart was discovered \citep[][]{Roberts1998,Roberts1999}
and extended TeV emission was observed $\sim8'$ west
of the PWN \citep[][]{hessrabbit2006,HESS2018}. The central engine of the PWN was later 
identified to be a relatively young gamma-ray pulsar \citep[PSR~J1418$-$6058;][]{Abdo2009}
(J1418 hereafter)
with a characteristic age of $\tau_{c}\approx10^4$\,yr,
and 
X-ray pulsations were suggested\footnote{https://cxc.harvard.edu/cdo/snr09/pres/Roberts\_Mallory.pdf}
and later confirmed \citep[][]{Kim2020}; in this work the pulsar's X-ray spectrum was
fitted by a power law with a hard photon index $\Gamma=1.0\pm0.6$.
Distance to the PWN was estimated to be 3.5--5.6\,kpc 
based on a column density and ISM morphology study \citep[][]{Voisin2019}.

	\citet[][]{Kishishita2012} carried out an X-ray study of
the Rabbit PWN using Suzaku data and found that the PWN spectra extracted from annular regions
soften with increasing distance from the pulsar, indicating significant synchrotron cooling
in the Rabbit PWN as has been observed in young PWNe \citep[e.g.,][]{nhra+14,amrk+14}.
However, a high-resolution Chandra image (e.g., Fig.~\ref{fig:fig1}) found 
a handful of point sources and a torus or jet-like structure around the pulsar \citep[][]{Ng2005,Kim2020}.
Since these features as well as the (hard) pulsar emission were not
resolved in the Suzaku data, their X-ray emission 
has likely biased the Suzaku  analysis,
making the X-ray spectra appear hard in the inner regions where
the contamination is more severe. Hence, revisiting a SED study of the Rabbit PWN by scrutinizing the contaminating sources is warranted. 

	In this paper, we present a 
	recently acquired NuSTAR observation and use archival Chandra and XMM-Newton data
to reveal the X-ray emission properties of the Rabbit PWN, independently of its pulsar
and contaminating sources.
We present our data reduction
in Section~\ref{sec:sec2_1}, and show 
detailed timing analysis
in Section~\ref{sec:sec2_2}.
We determine X-ray spectra of the pulsar and other contaminating sources within the PWN, and
assess their impact on measuring the PWN emission in Section~\ref{sec:sec2_3}.
We then carry out an imaging analysis (Section~\ref{sec:sec2_4}), and investigate spatially-integrated
and resolved the emission of the PWN
in Section~\ref{sec:sec2_5}.
We inspect data taken by the Fermi Large Area Telescope \citep[LAT;][]{fermimission}
to confirm the GeV measurements reported in the 4FGL DR-2
catalog \citep[{\tt gll\_psc\_v27.fit};][]{fermi4fgl} in Section~\ref{sec:sec3}.
After collecting multiband spectral data,  we modeled the broadband SED and X-ray spectral variation of the PWN (Section~\ref{sec:sec4}). We discuss the results from our model fitting and infer the physical properties of the PWN in Section~\ref{sec:sec5}. 
Note that all errors are at the 1$\sigma$ level and quoted flux values are absorption-corrected ones throughout the paper.
 
\section{X-ray Data Analysis}
\label{sec:sec2}

\subsection{Data reduction}
\label{sec:sec2_1}
       We use the 70-ks Chandra and 120-ks XMM-Newton archival data taken on 2007 June 14 (Obs. ID 7640)
and on 2009 February 21 (Obs. ID 0555700101), respectively, and
hard X-ray data obtained through the NuSTAR campaign of TeV PWNe \citep[][]{Mori2021}
on 2021 April 20 for 140\,ks. 

	We processed the Chandra data using {\tt chandra\_repro}
of CIAO~4.13 along with the most recent calibration database (version 4.9.6).
The Chandra data are useful to identify and characterize contaminating
sources within the PWN, despite its limited coverage of the PWN due to the chip gaps (Fig.~\ref{fig:fig1} a).
The XMM-Newton MOS data were processed with the {\tt emproc} task of SAS~20211130\_0941 along with the most recent calibration database (updated in 2021 November).
Note that we analyzed only
the MOS data because the PN exposure did not cover the PWN and those data were already analyzed
for the pulsar \citep[][]{Kim2020}.
The XMM-Newton MOS data were further cleaned following the standard flare-removal
procedure.\footnote{https://www.cosMOS.esa.int/web/xmm-newton/sas-thread-epic-filterbackground}
We processed the NuSTAR data with {\tt nupipeline} integrated in HEASOFT v6.29 (CALDB 20211020)
using the {\tt SAA\_MODE=strict} flag as recommended by the NuSTAR
science operation center. Net exposures after this initial reduction are
70\,ks, 100\,ks, and 55\,ks for Chandra, XMM-Newton, and NuSTAR, respectively.
X-ray images of the Rabbit PWN and its surrounding regions are
displayed in Figure~\ref{fig:fig1}.

\subsection{NuSTAR Timing Analysis}
\label{sec:sec2_2}
	The 110-ms X-ray pulsations of J1418
were detected in the XMM-Newton PN data, but the detection significance was not very
high with a chance probability $p\approx 10^{-7}$ \citep[][]{Kim2020}.
Hence, independent confirmation of the X-ray pulsations would be
very useful. The high-energy sensitivity and superb timing resolution of
NuSTAR \citep[][]{hcc+13}
are beneficial for the detection of
pulsations of this rapidly spinning 
pulsar, characterized by a hard X-ray spectrum.

\begin{figure}
\centering
\includegraphics[width=3.15 in]{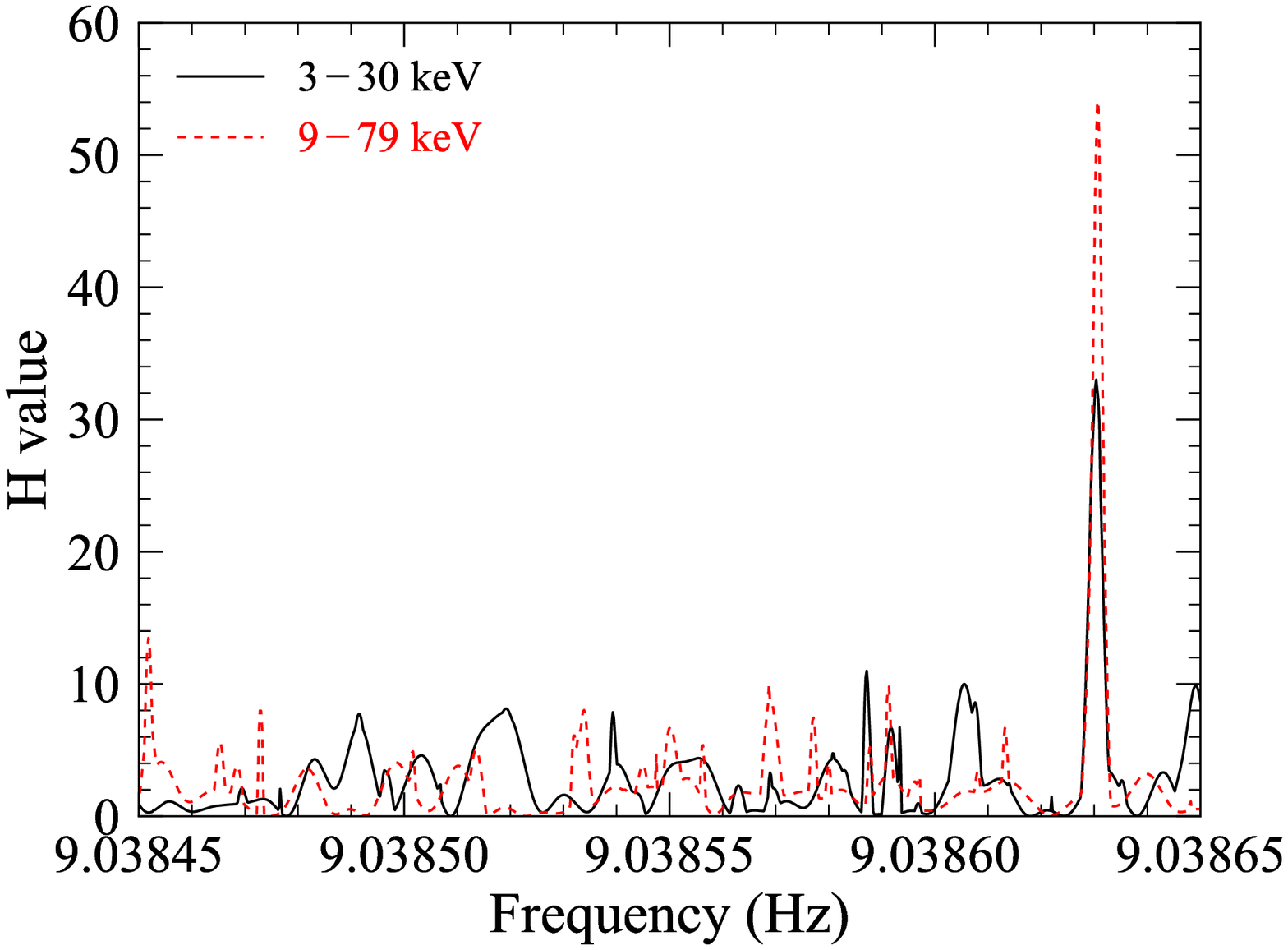} \\
\hspace{-4mm}
\includegraphics[width=3.05 in]{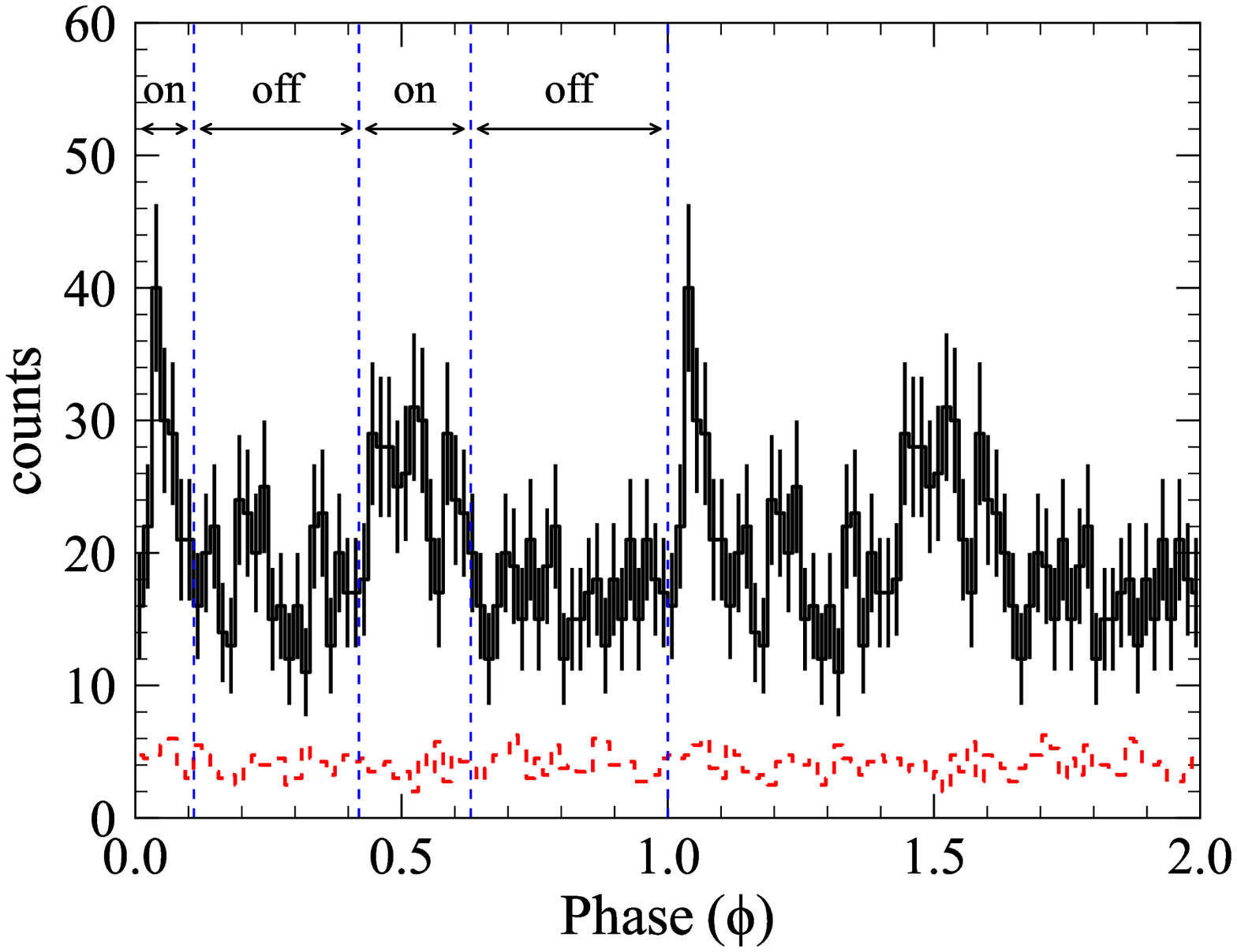} \\
\figcaption{{\it Top}: $H$-test results in the 3--30\,keV (black) and
the 9--79\,keV band (red).
{\it Bottom}: A 3--30\,keV pulse profile constructed by folding the source events
within $R=30''$ circles (FPMA and FPMB combined) on the best frequency of 9.03863\,Hz.
A folded background light curve is presented in red, and blue vertical lines
show on- and off-pulse intervals. \label{fig:fig2} }
\vspace{0mm}
\end{figure}

	We inspected the NuSTAR hard-band images (10--30\,keV) and identified
a point source in each of the FPMA and FPMB images. Assuming that the point source
is J1418 (see also Section~\ref{sec:sec2_4_1}),
we extracted source events within a $R=30''$ circle centered at the point source,
applied a barycenter correction to the event arrival times using the
pulsar position (R.A., decl.)=($214.677695^\circ$, $-60.967483^\circ$),
and performed timing analysis employing an $H$ test \citep[][]{drs89}
in the 3--30\,keV band. Extrapolating the pulsar timing solution \citep[][]{krjs+15}
derived from the Fermi-LAT data
predicts a spin frequency $f=9.03857$\,Hz
at the epoch of the NuSTAR observation.
However, it is possible that the extrapolated frequency may not be accurate due to large timing
noise and (undetected)  glitches.\footnote{https://www.slac.stanford.edu/$\sim$kerrm/fermi\_pulsar\_timing/}
Hence, we searched for pulsations in a broad range
around the extrapolated spin frequency ($f=9.03845$--$9.03865$\,Hz)
after fixing the frequency derivative $\dot f$ to the LAT-measured value of
$-$$1.383\times 10^{-11}\rm \ s^{-2}$. We detected significant pulsations
at $f=9.03863$\,Hz on MJD~59328 with an $H$ statistic of $H=33$, corresponding
to $p=4\times 10^{-5}$ (Fig.~\ref{fig:fig2} top)
after considering a trial factor of 22 (i.e. the number of independent frequency bins).
The significance increases slightly ($H=37$) in the 3--79\,keV band.
Due to the harder X-ray emission of the pulsar
(relative to other X-ray sources including the PWN),
we found a more significant detection of the pulsation 
in higher energy bands, e.g., 9--30\,keV and 9--79\,keV, with
$H=45$ and $H=54$, respectively. The latter corresponds to $p=9\times 10^{-9}$
(red curve in Fig.~\ref{fig:fig2} top).

	The pulse profile (Fig.~\ref{fig:fig2} bottom) is well measured with the NuSTAR data
-- a sharp spike and a broad bump with a phase separation of $\sim$0.5 are apparent.
Note that the spike ($\phi=0.05$) was seen, but the broad bump ($\phi\sim0.5$) was not well detected
in the previous XMM-Newton profile \citep[e.g.,][]{Kim2020}.
The NuSTAR profile allows a selection of on- ($\Delta \phi=$0--0.11 and 0.42--0.63) and
off-pulse phase intervals for pulsar and PWN studies, respectively.
Small features are also visible in the profile (e.g., $\phi\sim 0.2$), but they
disappear when using different energy bands or regions and thus are likely
caused by statistical fluctuations.

\subsection{Assessment of the point source contamination}
\label{sec:sec2_3}
	To accurately measure the spectrum of the PWN in the broad X-ray band
with the Chandra, XMM-Newton and NuSTAR data, we need to adequately account for
the contamination by point sources in the PWN. 
Although the pulsar contamination can be reduced in phase-resolved NuSTAR spectra 
by selecting the off-pulse phases, constant (off-pulse) pulsar emission may still be present. 
Furthermore, other X-ray sources, as seen in the Chandra and the XMM-Newton images (Fig.~\ref{fig:fig1}),
may affect the NuSTAR spectral analysis. 
Here we assessed contamination by the point sources in the NuSTAR's PWN spectrum,
and then characterized the
broadband X-ray spectra of the PWN (Section~\ref{sec:sec2_5}).

\newcommand{\marka}{\tablenotemark{a}}
\newcommand{\markb}{\tablenotemark{b}}
\newcommand{\markc}{\tablenotemark{c}}
\newcommand{\markd}{\tablenotemark{d}}
\begin{table*}[t]
\vspace{-0.0in}
\begin{center}
\caption{Power-law fit results for point sources in the PWN and for the PWN}
\label{ta:ta1}
\vspace{-0.05in}
\scriptsize{
\begin{tabular}{lccccccc} \hline\hline
data       & Instrument\marka  & energy range   & $N_{\rm H}$      & $\Gamma$  &  $F_{\rm 2-10keV}$ & $\chi^2$/dof\\
           &        & (keV)   & ($10^{22}\rm \ cm^{-2}$) &  & ($10^{-13}\rm \ erg\ s^{-1}\ cm^{-2}$) &   & \\ \hline
PSR pulsed & NuSTAR    & 3--30   & 2.78\markb & $0.94\pm0.33$   & $2.61^{+0.94}_{-0.69}$   & 8/14 \\
PSR off-pulse  & CXO+NuSTAR    & 0.5--10 & 2.78\markb & $2.21\pm0.51$   & $0.30^{+0.11}_{-0.08}$   & 27/43 \\
X1 & XMM  & 0.5--10 & $1.70\pm0.32$ & $1.02\pm0.17$ & $3.27^{+0.20}_{-0.19}$   & 45/40 \\ 
X2   & XMM  & 0.5--10 & $0.45\pm0.04$ & $3.05\pm0.13$ & $0.50\pm0.04$  & 71/63 \\
X2   & CXO  & 0.5--10 & $0.26\pm0.07$ & $2.35\pm0.16$ & $0.89^{+0.09}_{-0.08}$   & 24/23 \\ \hline
PWN\markc  & XMM      & 0.5--10 & $3.11\pm0.38\pm0.17$\markd & $2.16\pm0.14\pm0.07$\markd & $28.9\pm1.0\pm0.4$\markd &  324/324 \\ 
PWN\markc  & CXO      & 0.5--10 & $2.45\pm0.19\pm0.16$\markd & $1.78\pm0.12\pm0.05$\markd   & $35.4\pm1.0^{+2.0}_{-1.9}$\markd  &  137/164 \\ 
PWN\markc  & NuSTAR      & 5--20   & 2.78\markb & $2.05\pm0.08\pm0.04$\markd & $41.2\pm2.3\pm1.3$\markd   & 111/128 \\ 
PWN\markc  & CXO+XMM+NuSTAR  & 0.5--20 & $2.78\pm0.12$ & $2.02\pm0.05$ & $34.8\pm0.08$   & 688/642 \\ \hline
\end{tabular}}
\end{center}
\vspace{-0.5 mm}
\footnotesize{
$^{\rm a}${CXO: Chandra, XMM: XMM-Newton.}\\
$^{\rm b}${Fixed at the value measured from a joint fit of Chandra, XMM-Newton and NuSTAR data of the PWN.}\\
$^{\rm c}${Measured within a $R=3'$ circular region.}\\
$^{\rm d}${Second errors are systematic uncertainties estimated by varying background regions.}}
\end{table*}

\subsubsection{Contamination by the off-pulse emission of J1418}
\label{sec:sec2_3_1}
	X-ray spectra of the pulsar in the Chandra and XMM-Newton band (0.5--10\,keV)
were fit to a power-law model with 
$\Gamma\approx 1$ and $\Gamma\approx1.5$
for the pulsed and total emission, respectively,
in our previous XMM-Newton study \citep[][]{Kim2020}. We also determined the on-pulse spectrum of the pulsar
using the NuSTAR data. We extracted photon events within a $R=30''$ circle
in the on- and off-pulse data for the source and the background spectra, respectively.
Response files were computed with the {\tt nuproduct} tool.
After grouping the spectrum to have a minimum of 30 events per bin, we fit the 3--30\,keV
spectrum with an absorbed power-law model in XSPEC v12.12 holding the hydrogen column density
fixed at $N_{\rm H}=2.78\times 10^{22}\rm \ cm^{-2}$
(see Section~\ref{sec:sec2_5_2}).
Note that we used the {\tt tbabs} model
along with the {\tt vern} cross section \citep[][]{vfky96}
and {\tt angr} abundance \citep[][]{angr89} for the Galactic absorption throughout this paper
to compare with the previous Suzaku study of Rabbit \citep[][see Section~\ref{sec:sec5_1}]{Kishishita2012}\footnote{Using the newer {\tt wilm} abundance table, we obtain a larger $N_{\rm H}$ of $(4.14\pm0.18)\times 10^{22}\rm \ cm^{-2}$ for the $R=3'$ PWN (Section~\ref{sec:sec2_5_2}), but the other parameter values did not change significantly.}.
The best-fit parameters are $\Gamma=0.94\pm0.33$ and
2--10\,keV flux of $F_{\rm 2-10 keV}=2.61^{+0.94}_{-0.69}\times 10^{-13}\rm \ erg\ cm^{-2}\ s^{-1}$.
The latter corresponds to $8.36^{+3.00}_{-2.21}\times 10^{-14}\rm \ erg\ cm^{-2}\ s^{-1}$ when averaged over a spin cycle.

	We next estimated the off-pulse emission of the pulsar by analyzing
the Chandra data. We extracted a source spectrum using a $R=2''$ circular region
centered at the pulsar position, and a background spectrum was extracted from
a $R=2.1''$--$5''$ annular region around the pulsar.
Response files were generated with the {\tt specextract} tool.
The Chandra spectrum was grouped to have a minimum of 5 events per bin, and we used 
{\tt lstat}\footnote{https://heasarc.gsfc.nasa.gov/xanadu/xspec/manual/XSappe\\ndixStatistics.html}
\citep[][]{l92}
in XSPEC. We fit the spectrum with an absorbed power-law model and reproduced the earlier results
for the pulsar's total (pulsed + off-pulse) emission \citep[][]{Kim2020}.
To estimate the off-pulse spectrum, we fit the Chandra data with a blackbody plus
power-law or a double power-law model with the parameters of the second model component (power law)
held fixed at the best-fit values obtained for the pulsed spectrum in the NuSTAR analysis above.
Both provided an acceptable fit, but the blackbody plus power-law model
yielded an unreasonably high $kT=0.66\pm0.14$\,keV ($\chi^2$/dof=28/43)
for thermal emission from isolated neutron stars. 
The double power-law fit resulted in $\Gamma=2.21\pm0.51$ and
$F_{\rm 2-10 keV}=2.98^{+1.10}_{-0.80}\times 10^{-14}\rm \ erg\ cm^{-2}\ s^{-1}$ ($\chi^2$/dof=27/43)
for the first power-law component (with frozen
$N_{\rm H}=2.78\times 10^{22}\rm \ cm^{-2}$). Additional uncertainties due to the on-pulse spectral model
were estimated to be $\Delta \Gamma=^{+0.81}_{-0.57}$ and
$\Delta F_{\rm 2-10 keV}=^{+1.71}_{-1.31} \times 10^{-14}\rm \ erg\ cm^{-2}\ s^{-1}$.
Although the large error bars associated with the pulsed spectral parameters
and cross-calibration uncertainties of Chandra and NuSTAR \citep[][]{mhma+15}
do not allow a precise estimation of the off-pulse emission, its soft and
faint spectrum seems not to significantly contaminate
the NuSTAR PWN spectrum at $>$3\,keV (see Table~\ref{ta:ta1}).
We further investigate below (Section~\ref{sec:sec2_5_2}) any contamination by this off-pulse emission
in the NuSTAR analysis.

\subsubsection{Contamination by other sources}
\label{sec:sec2_3_2}
	In the high-resolution Chandra data, we identified several contaminating
sources within the PWN (e.g., S1--S8 within the $R=3'$ circle shown in Fig.~\ref{fig:fig1} a).
In addition, another variable source (X1)
in the north of S1 appeared only in the XMM-Newton data (Fig.~\ref{fig:fig1} b) as reported
in our previous study \citep[][]{Kim2020}.
While S1--S8 are quite faint, X1 was brighter than the pulsar in the XMM-Newton observation. 
Note, however, that X1 is highly variable; it was very faint or undetected in other Chandra/XMM-Newton
and our NuSTAR observations (Section~\ref{sec:sec2_4_1}).
Note also that there is another bright source (X2 in Fig.~\ref{fig:fig1} b)
at $\gapp3'$ east of J1418. This source
was also seen in the Suzaku data and can affect our spectral analysis of $R>3'$ regions.

	We first measured an X-ray spectrum of X1 using the XMM-Newton data.
We extracted the source spectrum within a $R=10''$ circle and grouped
it to have at least 30 counts per spectral bin. 
A background spectrum was extracted from a $R=20''$ circle
within the PWN in order to properly account for the PWN background
within the source region. Response files for the point source were produced
with the {\tt rmfgen} and {\tt arfgen} tasks of SAS.
We fit the XMM-Newton EPIC spectra with an absorbed power-law model.
The fit was acceptable with $\chi^2$/dof=45/40 ($p$=0.26),
yielding the best-fit parameters of 
$N_{\rm H}=(1.70\pm0.32)\times 10^{22}\rm \ cm^{-2}$,
$\Gamma=1.02\pm0.17$, and $F_{\rm 2-10keV}=3.27^{+0.20}_{-0.19}\times 10^{-13}\rm \ erg\ cm^{-2}\ s^{-1}$
(Table~\ref{ta:ta1}).
A single blackbody model fit was also acceptable with $kT=1.72\pm0.12$\,keV and
$N_{\rm H}=(0.37\pm0.20)\times 10^{22}\rm \ cm^{-2}$ ($\chi^2$/dof=48/40)
without requiring an additional component
with an $F$-test probability of $\approx 0.2$.
Below we assumed the power-law spectrum for X1 as a conservative estimate for
the hard X-ray band contamination (Section~\ref{sec:sec2_5_2}).

X2 was detected in both the XMM-Newton and Chandra data.
For the XMM-Newton data analysis, we extracted source and background spectra
using $R=16''$ and $R=32''$ circles, respectively.
We grouped the source spectrum to have at least 30 events per bin,
fit the spectrum with a power-law model, and inferred the best-fit parameters of
$N_{\rm H}=(4.52\pm0.40)\times 10^{21}\rm \ cm^{-2}$, $\Gamma=3.05\pm0.13$, and
$F_{\rm 2-10 keV}=(4.98^{+0.44}_{-0.40})\times 10^{-14}\rm \ erg\ cm^{-2}\ s^{-1}$.
We also analyzed the Chandra data using a $R=3''$ circle and a $R=3.1''$--$6''$ annulus
for the source and background spectrum, respectively. A power-law fit to the Chandra spectrum resulted in
$N_{\rm H}=(2.55\pm0.74)\times 10^{21}\rm \ cm^{22}$, $\Gamma=2.35\pm0.16$, and
$F_{\rm 2-10 keV}=(8.86^{+0.87}_{-0.79})\times 10^{-14}\rm \ erg\ cm^{-2}\ s^{-1}$ (Table~\ref{ta:ta1}).
The Chandra and XMM-Newton results are significantly different, meaning that
this source is variable. The source is outside the $R=3'$ circle, and its emission
is weak and spectrally soft, hence its influence on the NuSTAR analysis should not be substantial.

We analyzed Chandra spectra for S1--S8 extracted from $R=2''$
circles around the source positions, collecting 20--60 counts for each source.
Although their spectral parameters ($N_{\rm H}$ and $\Gamma$) are not well constrained due to the paucity of counts,
contamination from these faint sources to the NuSTAR PWN spectra above 3\,keV can be ignored. We verify this in Section~\ref{sec:sec2_5_2} using the Chandra data.

\subsection{Image analysis}
\label{sec:sec2_4}

\begin{figure}
\centering
\hspace{-3.0 mm}
\includegraphics[width=3.2 in]{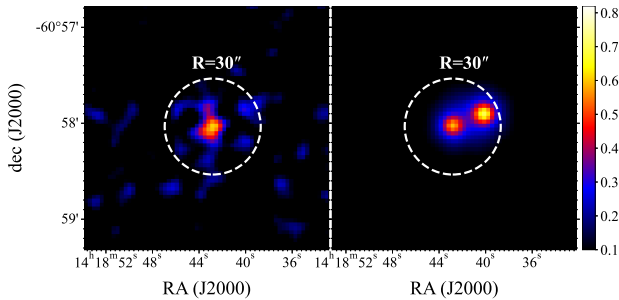}
\figcaption{An observed `on$-$off' image (left) and a simulated image of J1418+X1 (right).
{\it Left}: An exposure-scaled off-pulse image is subtracted from an on-pulse one in the 10--30\,keV band.
{\it Right}: A simulated image of the pulsar (center) and X1 (west) assuming that
X1 was as bright as was seen in the XMM-Newton data.
X1 is placed at $\sim$21$''$ northwest of J1418 (as observed by XMM-Newton).
$R=30''$ white circles are shown for reference, and the figures are smoothed, and scales are adjusted for better legibility.
\label{fig:fig3}
}
\vspace{0mm}
\end{figure}

\subsubsection{Pulsar image in the NuSTAR data}
\label{sec:sec2_4_1}
	We first inspected the NuSTAR image of the central region.
This is particularly important since the variable source X1, which
had a hard spectrum (see Section~\ref{sec:sec2_3_2})
and was brighter than the pulsar in the XMM-Newton data,
is only $\sim$20$''$ west of the pulsar (Fig.~\ref{fig:fig1} b) and
might have been bright during the NuSTAR observation, contaminating the PWN emission.
The detection of the pulsations from J1418 (Section~\ref{sec:sec2_2}) in the NuSTAR data already
suggests that X1 was not bright during the NuSTAR observation.

	In order to identify hard X-ray point sources (e.g., pulsar and X1),
we inspected on- and off-pulse NuSTAR images in the 10--30\,keV band and
clearly detected only one point source in each of FPMA and FPMB in the on-pulse intervals.
We aligned FPMA and FPMB images using the point source positions and produced 
a 10--30\,keV `on$-$off' image of the source (Fig.~\ref{fig:fig3} left).
Assuming that X1 is as bright
as it was in the XMM-Newton observation, we investigated  the effect of X1 using an image simulation.
We estimated 10--30\,keV count rates of J1418 and X1 using the measured spectra (Table~\ref{ta:ta1}),
and simulated a NuSTAR image by convolving the two point sources (offset by $21''$ with each other) with the NuSTAR PSF. 
A simulated image is displayed in the right panel of  Figure~\ref{fig:fig3}. 
Our simulation shows that X1 should have been resolved from the pulsar in the NuSTAR images, 
noting that NuSTAR's FWHM is 18$''$ \citep[e.g.,][]{amwb+14}.
Comparing the observed (left) and the simulated (right) images suggests that X1 was not bright during the NuSTAR observation. We further investigate the impact of (fainter) X1 emission in the spectral analysis below (Section~\ref{sec:sec2_5_2}).

\subsubsection{NuSTAR PWN image analysis} 
\label{sec:sec2_4_2}

	We used the off-pulse NuSTAR data to produce a PWN image in the 3--20\,keV band
because the background dominates over the PWN emission above 20\,keV.
The NuSTAR background is dominated by a non-uniform aperture component which produces
chip-to-chip variation.
We simulated background images using the {\tt nuskybgd}\footnote{https://github.com/NuSTAR/nuskybgd} tool
to account for the aperture background component. 
While the simulated images reflect well the aperture and the detector background components,  we found 
that the simulated counts for each chip differ from the observed ones by 16\%,  16\%, and $-24$\% 
for detector chips 1, 2, and 3, respectively.
We renormalized the simulated background level to match the observed background counts by adjusting the {\tt usernorm} parameters for the aperture, CXB, and GRXE components in {\tt nuskybgd}; after this process, the differences between the observational and simulated counts in the background regions were $\le$1\%.
Then the simulated backgrounds were subtracted from the observed images.
The resulting 3--20\,keV image (FPMA and FPMB combined) is displayed
in Figure~\ref{fig:fig1} c. The NuSTAR image also clearly shows extended X-ray emission which is well
contained within a $R=3'$ circle.

\begin{figure}
\centering
\includegraphics[width=3. in]{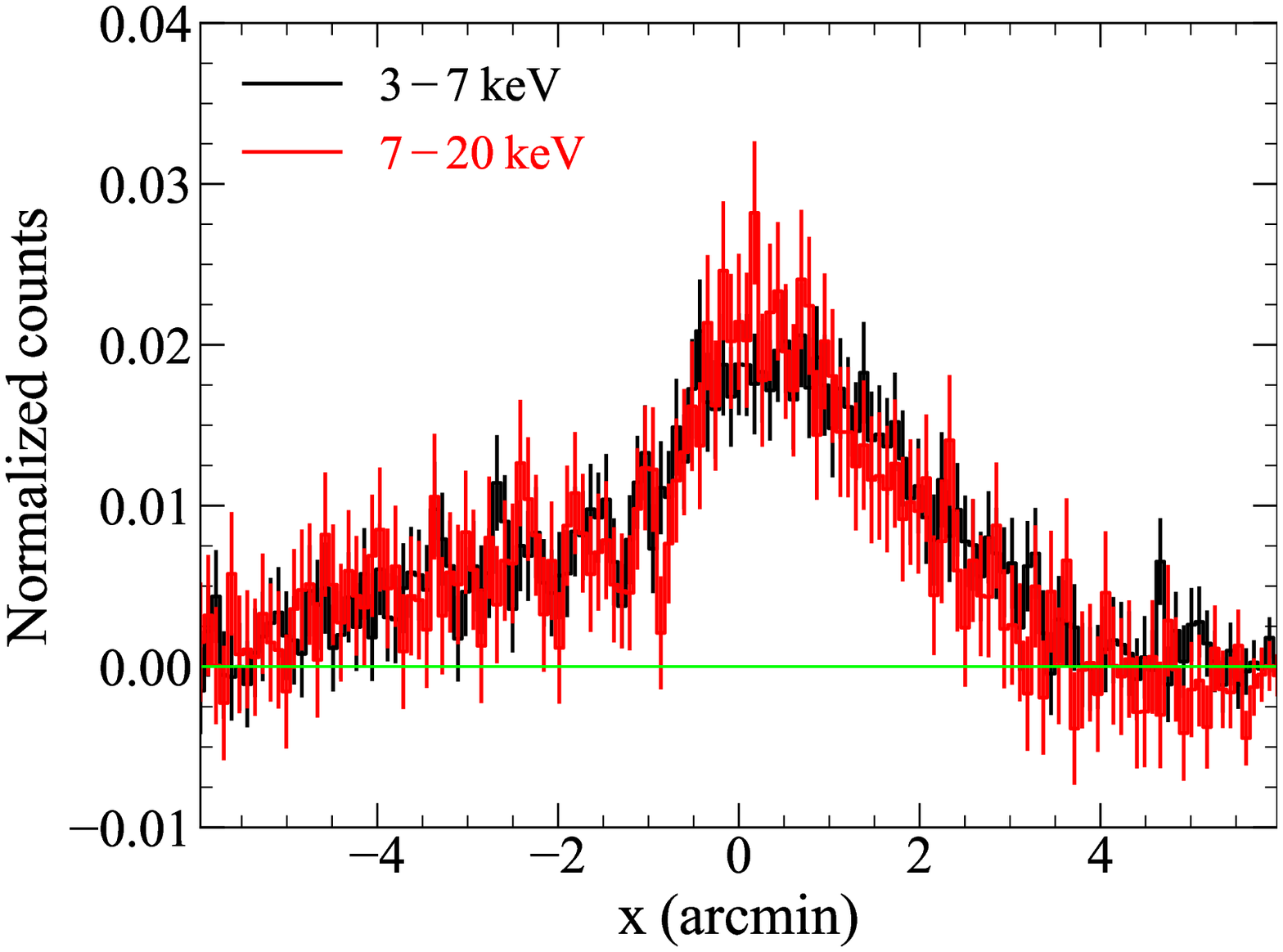} 
\includegraphics[width=3. in]{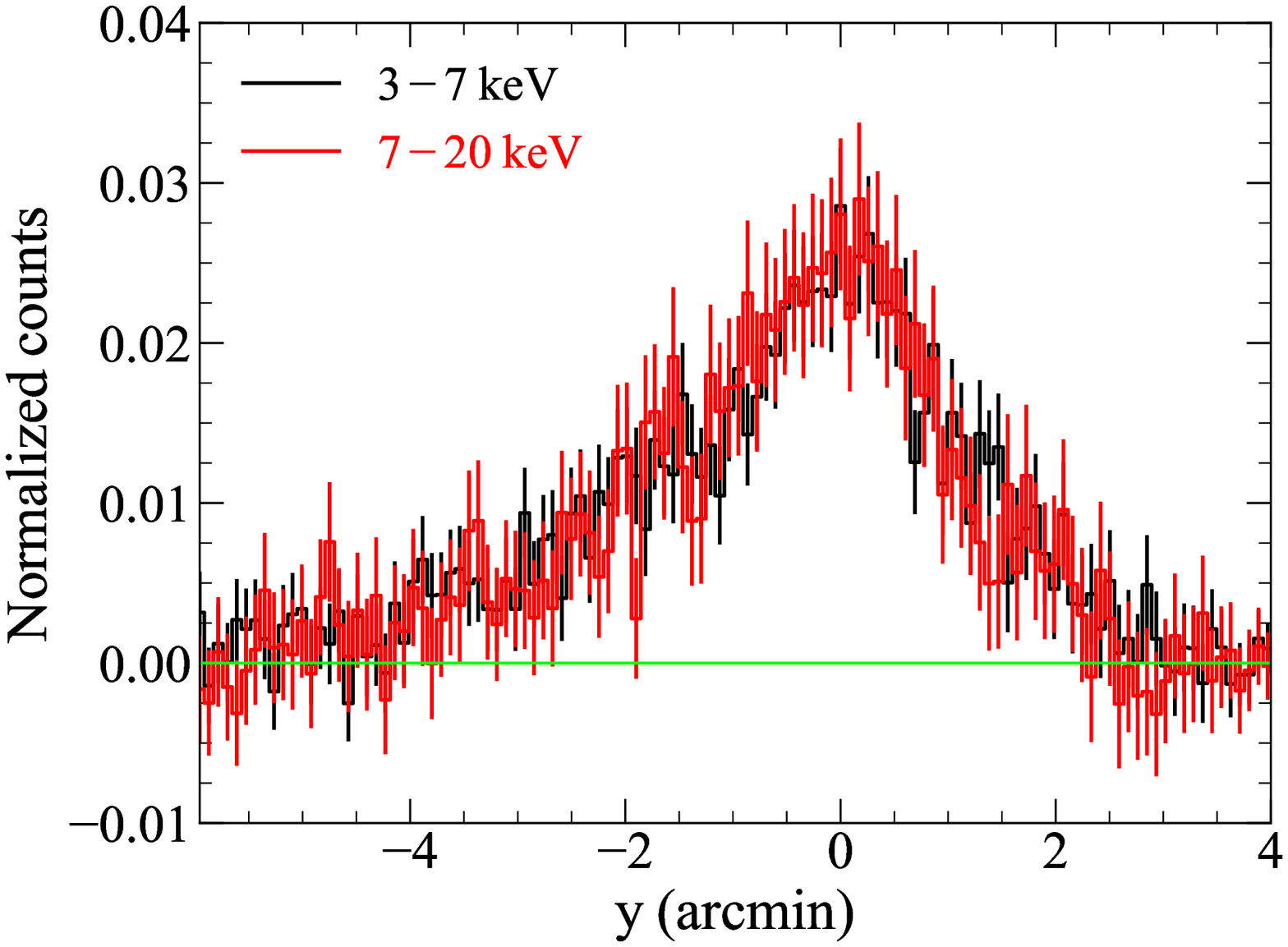}
\figcaption{3--7\,keV and 7--20\,keV NuSTAR profiles projected onto the x- and y-axis
defined in Figure~\ref{fig:fig1} c.\label{fig:fig4}}
\vspace{0mm}
\end{figure}

	Given the broadband NuSTAR data, we investigated whether the PWN size varies 
with energy. To capture the south-west extension,
we rotated the combined NuSTAR image
by $38^\circ$ from west to north with the origin at the pulsar position
(Fig.~\ref{fig:fig1} c), and defined the horizontal and vertical directions as
the x and y axes, respectively. We then selected  two energy bands 
in which the numbers of source counts are approximately the same.
We adopted a $12' \times 10'$ box and projected the box image onto the x and y axes.
The projected profiles are displayed in Figure~\ref{fig:fig4}.
The low- and high-energy profiles projected onto the x-axis display a small difference;
the latter appears to be slightly narrower than the former (Fig.~\ref{fig:fig4} top),
possibly indicating an energy-dependent size  shrinkage in the southwest (tail) direction.
However, the difference is not statistically significant, and we conclude 
that X-ray spectral softening is not observed in the Rabbit PWN with the current observation.

\subsubsection{XMM-Newton and Chandra PWN images on large scales}
\label{sec:sec2_4_3}
	To better identify the extended emission of the PWN, we produced images on large scales using
the XMM-Newton and Chandra data. For this, we used filter wheel closed (FWC) data and blank-sky data
for the XMM-Newton and Chandra analysis, respectively, to adequately remove the particle-induced background.

	For the XMM-Newton data analysis, we generated the filter wheel closed (FWC) data using
the {\tt evqpb} task of SAS. The FWC image appeared similar to the observed one in
source-free regions, but some (energy-dependent) differences were noticeable, especially at low energies.
In addition, the exposure-time ratio of the observational and FWC data was different from the measured
count ratios in the source-free regions perhaps because of the temporal variation of particle flares.
To derive a normalization factor appropriate for the measured count ratios,
we compared the observational and FWC data taken outside the field of view (FoV).
The spectral shapes of the observational and FWC data agreed well at $\gapp$1--2\,keV.
By comparing the $>$2\,keV spectra, we derived normalization
factors of 0.45 and 0.33 for MOS1 and MOS2, respectively, which are smaller
than the exposure ratio of 0.48 (see also Section~\ref{sec:sec2_5}).
We then generated images of the full FoV, subtracted the FWC image from the observed one,
and divided the FWC-subtracted image by the exposure map. A MOS1+MOS2 image in the 2--8\,keV band
is displayed in Figure~\ref{fig:fig1} b. The source is extended primarily in the NE-SW direction
and the bright tail emission is contained within a $R\sim 3'$ region.
There appears some emission slightly outside the $R=5'$ circle (north and northwest).
While this may be the PWN emission, we found that the brightness profiles
in the north and north-west directions do not follow a monotonic trend;
the brightness decreases to 4--5$'$ and then increases at $R\gapp5'$.
This outer emission might have been produced
by some other sources or by imperfect subtraction of the FWC background near
the chip boundaries \citep[e.g.,][]{Kuntz2008}.

	We adopted the above procedure for the Chandra data analysis using the blank-sky data.
We generated the blank-sky events with the {\tt blanksky} task of CIAO.
For the Chandra data, we found that the exposure-time ratio adequately
explains the measured count ratios of the observational and blank-sky data
in low-brightness regions of the former.
We produced observational and blank-sky images in the 2--7\,keV band,
subtracted the latter from the former, and divided the resulting image by the exposure map.
The image is displayed in Figure~\ref{fig:fig1} a.
The Chandra image resembles the XMM-Newton one, and we found that the brightness in the north and
north-west directions show a non-monotonic trend as was seen in the XMM-Newton image.

\subsection{Spectral analysis}
\label{sec:sec2_5}
In this section, we measure the spatially integrated and resolved spectra of the
Rabbit PWN, taking into account the point source contamination.

\subsubsection{Background spectra for the extended PWN}
\label{sec:sec2_5_1}
	Because the source emission extends to large distances from J1418
(Fig.~\ref{fig:fig1}), we need to extract background spectra from source-free
regions far away from the PWN. In this case, the detector
and particle-induced backgrounds may not represent well the source-region background.
To mitigate this, we used the FWC and blank-sky data for the XMM-Newton and Chandra analysis,
respectively, as was done for the image analysis.
For NuSTAR data analysis, we employed the {\tt nuskybgd} simulations \citep[][]{Wik2014}.

For the XMM-Newton data, we selected background regions at $>7'$ and collected spectra from the
observational and FWC data. These spectra showed a prominent instrumental line at $\sim$1.5\,keV. We compared
the continuum-subtracted lines measured in the observational and FWC data,
and verified the normalization factors obtained by comparing the out-of-FoV continuum spectra
(Section~\ref{sec:sec2_4_3}).
We subtracted the FWC spectrum from the observed one to remove the particle-induced background
and produced a `sky background' spectrum.
We did the same for the source-region spectra using the same normalization factors obtained above and
constructed the source spectra. The `sky background' spectrum collected
far away from the source region would be an underestimation of the source-region background due to
the optics vignetting effect. To correct for it, we multiplied the background spectrum by the energy-dependent
effective area (i.e., ancillary response files; ARF) ratio of the source and background regions. In these processes,
we carefully propagated the counting uncertainties.

\begin{figure}
\centering
\includegraphics[width=3. in]{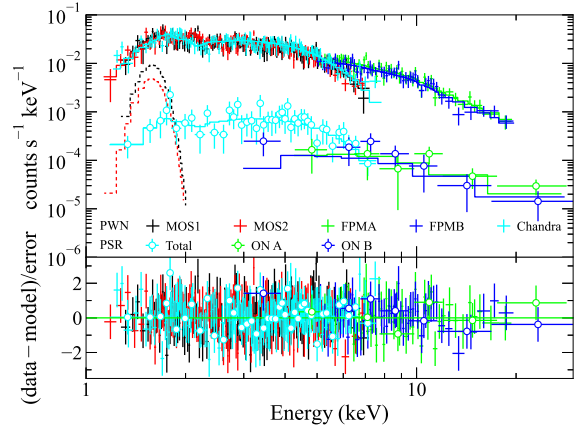} 
\figcaption{X-ray spectra of the PWN (crosses; 1--20\,keV) extracted within
a $R=3'$ circle and of the pulsar (empty circles; 1--30\,keV).
The best-fit instrumental Gaussian lines for the PWN spectra measured by XMM-Newton
are presented in dashed black (MOS1) and red (MOS2) lines in the top panel.
The pulsar spectrum measured by Chandra (cyan circles in the top panel)
includes both the pulsed and off-pulse components, and
those by the NuSTAR are only for the pulsed component (blue and green circles).
The bottom panel shows residuals after subtracting the best-fit power-law and instrumental Gaussian models. \label{fig:fig5}}
\vspace{0mm}
\end{figure}

The XMM-Newton spectra constructed following the aforementioned procedure exhibited
small but noticeable residuals around the instrumental line complex at 1.5--2\,keV; the residuals (e.g., dashed black and red lines in Fig.~\ref{fig:fig5} top) show a structure that is broader than individual instrumental lines.
This is possibly caused by temporal variation of the line emissions and likely imperfect
subtraction of the FWC background \citep[e.g.,][]{Kuntz2008}.
Therefore, we added a broad Gaussian line along with diagonal response files to our spectral model (Section~\ref{sec:sec2_5_2}--\ref{sec:sec2_5_4});
the typical central line energy and width of the Gaussian were estimated to be
$\sim$1.5\,keV and $\sim$0.15\,keV, respectively.

A similar procedure was applied to the
Chandra analysis with the blank-sky data.
We constructed the source and background spectra within the PWN (see below) and the source-free regions of the observation data, respectively.
We then produced the blank-sky spectra within the same source and background regions
and subtracted the blank-sky emissions from the observed source and background spectra.
We checked that the blank-sky-subtracted spectra did not show any noticeable instrumental line.
We then scaled the background spectra by the effective area (ARF) ratio of the source and background regions. 
For the NuSTAR data analysis, we used the off-pulse data to construct the source
spectra and performed {\tt nuskybgd} simulations using annular regions at $R>5'$
from J1418 to generate the background spectra
appropriate for the source regions \citep[e.g.,][]{Wik2014}.

	Using this method, we measure the PWN spectra within a few representative regions, namely $R=3'$,
$R=5'$, and $R=7'$ circles. The latter two are to be compared with previous Suzaku and ASCA results.
We adopted an absorbed power-law model for our spectral analyses
and fit the data in the 0.5--10 keV (XMM-Newton and Chandra) and 5--20 keV bands (NuSTAR).

\subsubsection{PWN spectra of the bright $R=3'$ region}
\label{sec:sec2_5_2}
	To measure the PWN spectra within the $R=3'$ region (Fig.~\ref{fig:fig1}),
we excluded J1418, S1, S2, X1 (using a $R=40''$ circle), X2 (slightly outside the $R=3'$ circle),
and three faint sources (S4, S5, and S8; $R=16''$ circles) from the XMM-Newton data (Fig.~\ref{fig:fig1} a and b).
We excised nine point sources (S1--S8 and J1418; $R=2''$ circles) from the Chandra data but
did not excise any region from the NuSTAR off-pulse data.
We verified that the contamination by S1--S8 affects only the estimation of the flux ($\sim$4\%) not the spectral slope ($\Gamma$) by comparing the Chandra spectra measured with and without S1--S8.

We first fit the XMM-Newton, Chandra, and NuSTAR spectra separately
after grouping them to have at least 100, 50, and 50 counts per spectral bin, respectively.
From the XMM-Newton data, we inferred the best-fit parameters of
$N_{\rm H}=(3.11\pm0.38)\times10^{22}\rm \ cm^{-2}$, $\Gamma=2.16\pm0.14$, and
$F_{\rm 2-10\,keV}=(2.89\pm0.10)\times 10^{-12}\rm \ erg\ cm^{-2}\ s^{-1}$.
The fit was formally unacceptable with $\chi^2$/dof of 435/347,
and a residual trend at low energies ($<$1\,keV) was noticeable.
We verified that ignoring the low-energy data improved the fit ($\chi^2$/dof=324/324)
without altering the best-fit parameter values significantly.
Note that the best-fit values change depending on the background-region selection by
$\Delta N_{\rm H}=\pm 0.17\times10^{22}\rm \ cm^{-2}$, $\Delta \Gamma=\pm 0.07$ and thus
$\Delta F_{\rm 2-10\,keV}=^{+4.32}_{-4.25}\times 10^{-14}\rm \ erg\ cm^{-2}\ s^{-1}$ (standard deviations).
The changes of $N_{\rm H}$ and $\Gamma$ are due to their covariance, and
we find $\Delta \Gamma=\pm 0.02$ for frozen $N_{\rm H}$ ($=2.78\times 10^{22}\rm \ cm^{-2}$).

The best-fit parameters obtained from the Chandra data are $N_{\rm H}=(2.45\pm0.19)\times10^{22}\rm \ cm^{-2}$,
$\Gamma=1.78\pm0.12$, and $F_{2-10 keV}=(3.54\pm0.10)\times 10^{-12}\rm \ erg\ cm^{-2}\ s^{-1}$
($\chi^2$/dof=137/164). The parameter values vary
depending on the background selection by $\Delta N_{\rm H}=\pm 0.16 \times 10^{22}$ cm$^{-2}$, $\Delta\Gamma=\pm 0.05$,
and $\Delta F_{\rm 2-10\,keV}=^{+1.98}_{-1.88}\times 10^{-13}\rm \ erg\ cm^{-2}\ s^{-1}$ (standard deviations).
$\Delta \Gamma$ was estimated to be $\pm0.05$ for frozen $N_{\rm H}$.
The XMM-Newton and Chandra results are discrepant
because of different excision regions and parameter covariance ($N_{\rm H}$ vs. $\Gamma$);
by freezing $N_{\rm H}$ to a common value, we found consistent $\Gamma$ in the XMM-Newton and Chandra fits.

We used the {\tt nuskybgd} simulations for the NuSTAR data analysis.
Since the simulations may be less accurate below 5\,keV\footnote{https://github.com/NuSTAR/nuskybgd}
and background dominates above 20\,keV,
we fit the NuSTAR spectra in the 5--20\,keV band. The best-fit parameters were
estimated to be $\Gamma=2.05\pm0.08$ and
$F_{\rm 2-10keV}=(4.12\pm0.23)\times 10^{-12}\rm \ erg\ cm^{-2}\ s^{-1}$ for
frozen $N_{\rm H}=2.78\times 10^{22}\rm \ cm^{-2}$ ($\chi^2$/dof=111/128).
To assess the systematic effects of the background selection for the {\tt nuskybgd} simulations,
we tried different background regions. In this case, the best-fit parameter values
change by $\Delta\Gamma=\pm0.04$ and
$\Delta F_{\rm 2-10keV}=\pm 1.3\times 10^{-13}\rm \ erg\ cm^{-2}\ s^{-1}$.
We cross-checked these results by an analysis performed with `in-flight' backgrounds.
The best-fit $\Gamma$ value did not alter significantly
but the flux value varied by $\sim\pm$10\% depending on the background region.
As noted above, the NuSTAR-measured flux includes contamination from S1--S8.

Additional uncertainties can be introduced in the NuSTAR analysis by contamination from the pulsar (off-pulse) and X1.
We assessed their effects by simultaneously modeling the emission in
the NuSTAR off-pulse data fits.
We first varied the best-fit parameters of the pulsar's off-pulse spectrum (Section~\ref{sec:sec2_3})
within their 68\% confidence intervals, considering the uncertainties in the on-pulse spectral model
and the covariance between $\Gamma$ and $F_{\rm 2-10 keV}$.
We then fit the PWN spectra with two power laws, holding the parameters for the second power law (pulsar emission)
at the varied values, and found that the influence of the pulsar emission is
ignorable (e.g., $\Delta\Gamma=+0.03$ and $\Delta F_{\rm 2-10 keV}$=$-3.33\times10^{-14}\rm \ erg\ cm^{-2}\ s^{-1}$). We did the same for the X1 emission. The variable source was faint, but its brightness was not well measured at the NuSTAR epoch. Hence we assumed that X1 was 30\% as bright as it was at the XMM-Newton epoch.
In this cases, $\Delta \Gamma$ and $\Delta F_{\rm 2-10 keV}$ were estimated to be
+0.02 and $-$2.5\%, respectively, which are smaller than the statistical uncertainties.

To characterize the $R=3'$ PWN emission better,
we jointly fit the XMM-Newton, Chandra, and NuSTAR spectra,
and inferred the best-fit parameters to be $N_{\rm H}=(2.78\pm0.11)\times 10^{22}\rm \ cm^{-2}$,
$\Gamma=2.02\pm0.05$, and $F_{\rm 2-10keV}=(3.48\pm0.08)\times 10^{-12} \rm \ erg\ cm^{-2}\ s^{-1}$.
The fit was acceptable with $\chi^2$/dof of 688/642,
but improved significantly ($\chi^2$/dof=577/619)
without altering the best-fit parameter values
when we ignored the low-energy ($<$1\,keV) XMM-Newton data (Fig.~\ref{fig:fig5} left).
The cross-normalization factors (set to 1 for Chandra) were estimated
to be $0.82\pm0.02$ and $1.17\pm0.06$ for MOS1 and FPMA, respectively.
The central $R=40''$ region excised from the XMM-Newton data has
a spectrum that is well characterized by a power law with $\Gamma=1.96\pm0.08$ and
$F_{\rm 2-10 keV}=(6.79\pm0.27)\times 10^{-13}\rm \ erg\ cm^{-2}\ s^{-1}$
($\chi^2$/dof=46/54), containing $\sim$20\% flux of the $R=3'$ PWN.
Noting that larger regions were excised from the XMM-Newton data and the NuSTAR data
included point sources (S1--S8, the pulsar's off-pulse and putative X1 emission),
the estimated normalization factors seem reasonable.
Additionally, imperfect cross-calibration of the instruments would introduce
some uncertainties \citep[][]{mhma+15}; differences in the flux calibration would be
included in the cross-normalization factors, and differences in the spectral-slope calibration
would affect the fit-inferred $N_{\rm H}$.
In the spectral analyses below (Sections~\ref{sec:sec2_5_3} and \ref{sec:sec2_5_4}), we held $N_{\rm H}$ fixed at $2.78\times 10^{22}\rm \ cm^{-2}$.

\subsubsection{Spectra of the more extended diffuse regions}
\label{sec:sec2_5_3}
We next characterize the source emission within larger regions ($R=5'$ and $R=7'$ circles) using the XMM-Newton and Chandra data to compare with previous ASCA and Suzaku measurements \citep[][]{Roberts2001a,Kishishita2012}.
We excised J1418, S2 and X1 (central 40$''$ circle), X2 (24$''$), and
other point sources (e.g., S4, S5, and S8 using 16$''$ circles) from the XMM-Newton data.
We removed the point sources (e.g., X1, X2, and S1--S8) from the Chandra analysis using $R=2''$ apertures.
We collected events within $R=5'$ to construct the source spectra, grouped them to have $>$100 events per bin,
and jointly fit the Chandra and XMM-Newton data.
The best-fit parameters were inferred to be $\Gamma=2.11\pm0.05$ and
$F_{\rm 2-10keV}=(5.72\pm0.18)\times 10^{-12}\rm \ erg\ cm^{-2}\ s^{-2}$.
The MOS1 cross normalization factor was measured to be 0.78$\pm$0.03 with respect to that of Chandra;
this can be ascribed to the larger excision apertures (e.g., the central $R=40''$) used for the XMM-Newton analysis.
The 2--10\,keV flux we measured is still lower by 9\%
than the previous Suzaku measurement of $F_{\rm 2-10keV}=(6.27\pm0.13)\times 10^{-12}\rm \ erg\ cm^{-2}\ s^{-2}$.
We suspect that this difference is due to the inclusion of point sources in the Suzaku analysis.
Indeed, by using all the emissions within the aperture except for X2 as was done
for the Suzaku data analysis of \citet[][]{Kishishita2012}, we found
$\Gamma=2.02\pm0.05$ and $F_{\rm 2-10keV}=(6.14\pm0.18)\times 10^{-12}\rm \ erg\ cm^{-2}\ s^{-2}$.
The systematic uncertainties on the flux estimations due to the background selection
are 2\% and 9\% for XMM-Newton and Chandra, respectively, and our results agree with 
the Suzaku one at $<1\sigma$ levels.

We also measured the larger $R=7'$ region spectrum
to compare with the previous ASCA result \citep[][]{Roberts2001a}.
For the comparison, we included all sources within the aperture as was done for the ASCA data analysis.
The XMM-Newton+Chandra spectra were fit with a power law having $\Gamma=2.35\pm0.05$ and
$F_{\rm 2-10keV}=(7.39\pm0.24)\times10^{-12}\rm \ erg\ cm^{-2}\ s^{-1}$ for frozen
$N_{\rm H}=2.78\times 10^{22}\rm \ cm^{-2}$. The flux value is consistent with $F_{\rm 2-10keV}=(7.33\pm0.17)\times10^{-12}\rm \ erg\ cm^{-2}\ s^{-1}$
measured by ASCA.
Note that the exact region size for the ASCA flux was not reported
and the ASCA-inferred $N_{\rm H}$ was substantially smaller.

\subsubsection{Spatially-resolved spectrum of the PWN}
\label{sec:sec2_5_4}
	We searched for any spatial variation of the X-ray spectrum due to the synchrotron burn-off effect
even though the NuSTAR imaging analysis did not show strong evidence for it (Section~\ref{sec:sec2_4_2}).

\begin{figure*}
\centering
\begin{tabular}{cc}
\includegraphics[width=3.4 in]{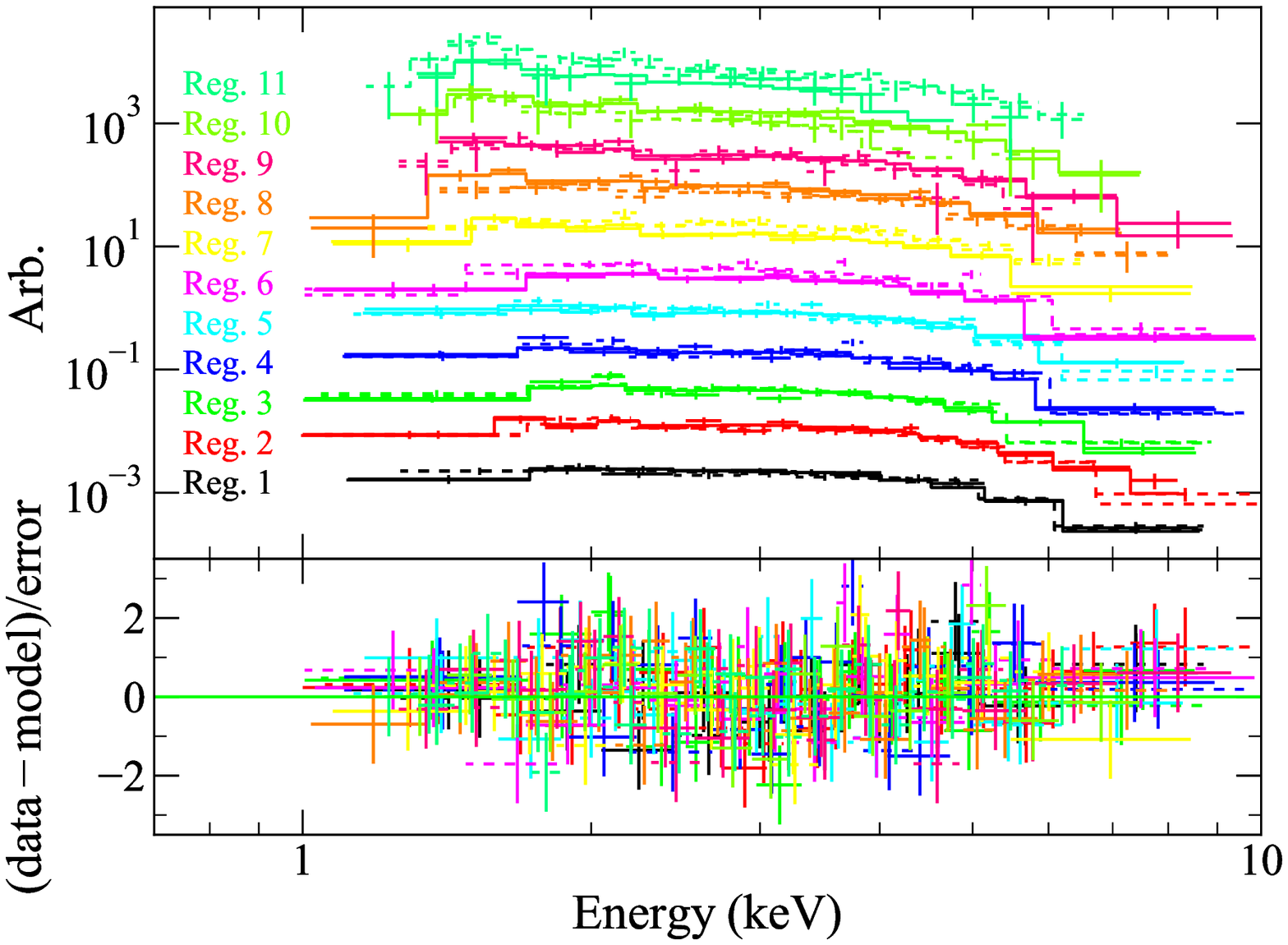} &
\includegraphics[width=3.5 in]{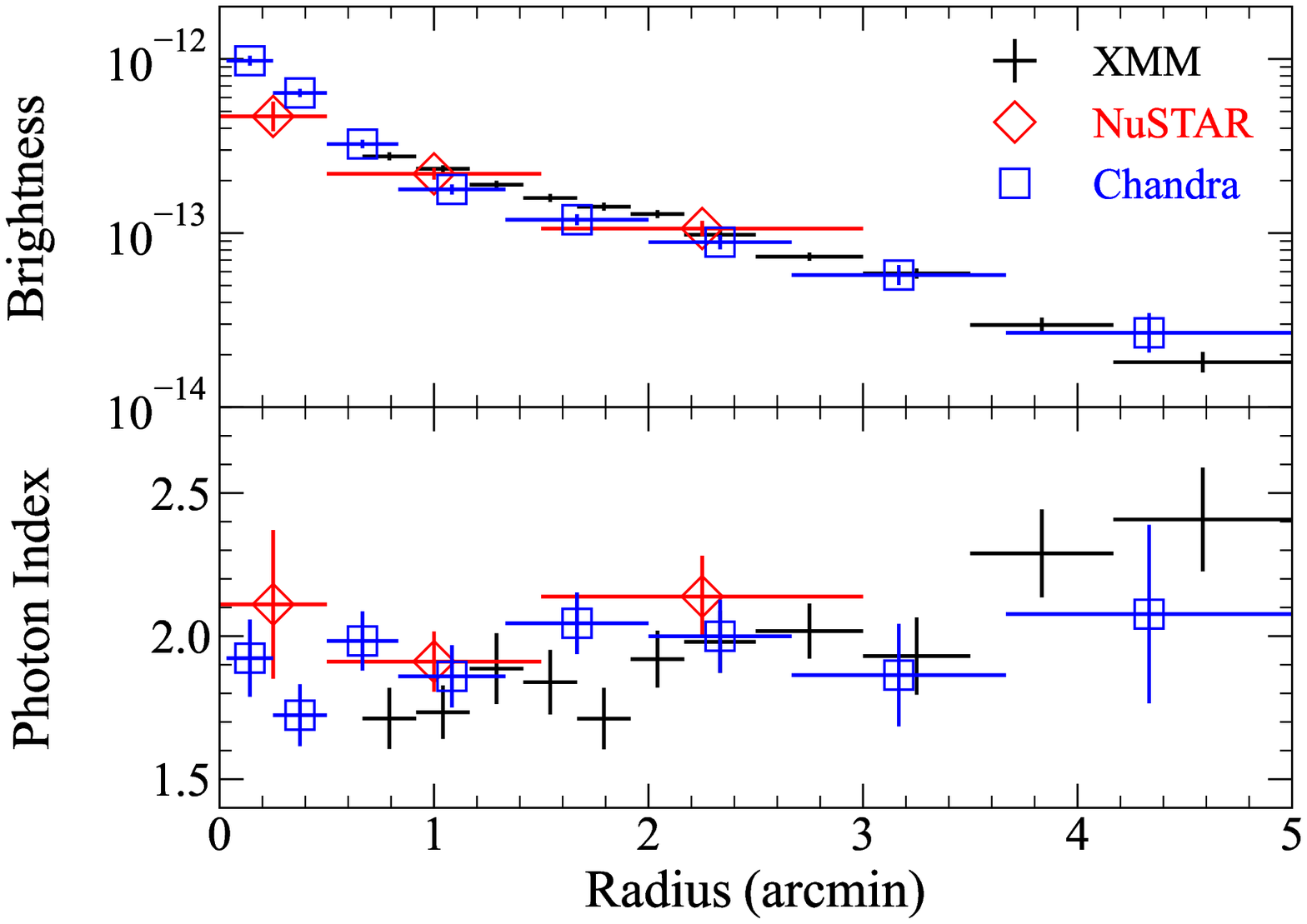} \\
\end{tabular}
\figcaption{Results of the spatially-resolved spectral analysis. Left: XMM-Newton spectra of the 11 annular regions. The spectrum of the innermost region (reg. 1) is shown in black, and the spectra of the outer regions are displaced for visibility. The MOS1 and MOS2 spectra are displayed as solid and dashed lines, respectively. Right: radial profiles of brightness (top) and photon index (bottom).
\label{fig:fig6}}
\vspace{0mm}
\end{figure*}

For the XMM-Newton data, we used 11 annular regions within $R=5'$.
These regions were selected to have different sizes
depending on the brightness. After excising the central $40''$ and point sources,
we constructed a spectrum for each region
and grouped the spectrum to have at least 50 counts per spectral bin.
We jointly fit the 11 spectra with an absorbed power law.
The other parameters were optimized separately for each region (i.e., untied $\Gamma$).
We also assessed systematic uncertainties due to background selection and
added them in quadrature to the statistical errors.
The results are displayed in Figure~\ref{fig:fig6}.
The brightness monotonically decreases
with increasing radius, and $\Gamma$ does not show a monotonic trend out to $\sim$3.5$'$ and then
slowly increases.
The fit was formally unacceptable with $\chi^2$/dof of 1751/1494 ($p\approx 4\times10^{-6}$), 
but ignoring low-energy data ($<$1\,keV) improved the fit ($\chi^2$/dof=1452/1366 and $p=0.05$) without altering the best-fit parameter values.
We compared the above results with ones obtained by a model with tied $\Gamma$ for the 11 spectra
and found that the untied $\Gamma$ model provides a better fit with an $F$-test probability of $3\times10^{-3}$.

In our further investigation, we found that the large $\chi^2$ was caused mostly by the low-energy spectra of outer regions. In addition,
the cross-normalization factors between MOS1 and MOS2 significantly deviated from 1 in outer regions;
the factors are consistent with 1 out to $R\approx 3.5'$, but they are $1.30\pm0.13$ and $1.69\pm0.22$
in the outermost two regions (Reg. 11 in Fig.~\ref{fig:fig6} left). These are probably because
the faint emission in the $R=3.5$--$5'$ regions is strongly affected
by the background and thus cannot be measured reliably. Furthermore, these regions are near the chip boundary where the particle-induced background is strong \citep[][]{Kuntz2008}, and
there seemed to be some contamination at $R>4'$ in the north and north-west directions
(Section~\ref{sec:sec2_4_3}).
Hence, we do not use the last two data points ($R\ge 3.5'$)
in Figure~\ref{fig:fig6} right for our modeling below. In this case, the untied $\Gamma$ model was not
significantly favored over the tied $\Gamma$ one with an $F$-test probability of 17\%.

We found similar results with the Chandra data (Fig.~\ref{fig:fig6} right).
The Chandra results agree well with the XMM-Newton ones but have larger uncertainties.
In the outermost region, Chandra data implied a smaller $\Gamma$, but it cannot
be discriminated from the XMM-Newton results due to the large uncertainty;
the faint emission spread over a large region was difficult to precisely characterize.

We also analyzed NuSTAR's off-pulse data to measure spatial variation
of the spectrum within a $R=3'$ region (within the FoV).
We extracted spectra using a $R=30''$ circle and two annular regions having widths of 60$''$ and 90$''$.
We grouped the NuSTAR spectra to have a minimum of 30 counts per bin and fit the 5--20\,keV
spectra with a power-law model having separate $\Gamma$ for each spectrum.
The fit was acceptable with $\chi^2$/dof=221/214, and
the results are presented in Figure~\ref{fig:fig6}.
The NuSTAR results broadly agree with the XMM-Newton and Chandra ones, but some difference is noticeable (e.g., the bottom panel of Fig.~\ref{fig:fig6} right) perhaps because of contamination from the point sources and cross-calibration issues.
We also tried to fit the data with a model having a common photon index
and found that the model provided an equally good fit ($\chi^2$/dof=224/216) with the best-fit photon index of $2.03\pm0.07$.  
An $F$-test comparison of the tied-$\Gamma$ and untied-$\Gamma$ models
gives $p\sim$0.24, implying insignificant spectral softening.

Note that our results are discrepant with the significant spectral softening
measured by Suzaku \citep[][]{Kishishita2012}. We speculated that this is because
the pulsar (and X1) emission was not removed in that analysis. To confirm, we analyzed
the XMM-Newton, Chandra, and NuSTAR data using the same annular regions as the Suzaku
ones without excising any point source.
Spectral softening trends, similar to the Suzaku measurement, were measured in
our analysis. In the innermost zone ($R<1'$), the Chandra and NuSTAR spectra were measured
to be consistent with the Suzaku spectrum ($\Gamma\approx 1.8$)
while the XMM-Newton spectrum is significantly harder ($\Gamma\approx 1.6$);
this is likely caused by contamination from X1.
In outer zones, the photon indices were measured
to be consistent with the Suzaku results ($\Gamma=2.1$--2.2).
Note again that in the outermost zone ($R>3'$), the cross normalization
factor between MOS1 and MOS2 significantly deviated from 1.

\section{Fermi-LAT Data Analysis}
\label{sec:sec3}

	We analyzed gamma-ray data taken with the Fermi LAT.
We extracted 100\,MeV--1\,TeV events acquired between 2008 August 4 and 2022 February 10
spanning approximately 13.5\,yrs.
The data were analyzed with Fermipy v1.0.1 \citep[][]{Wood2017} along with the {\tt P8R3\_SOURCE\_V3}
instrument response.\footnote{https://fermi.gsfc.nasa.gov/ssc}
We selected the {\tt Front+Back} event type in the {\tt SOURCE} class
within a $10^\circ\times 10^\circ$ square region of interest (RoI) centered
at 4FGL~J1417.7$-$6057 (LAT counterpart of Rabbit; J1417 hereafter) and
reduced the data using the zenith angle $<90^\circ$, DATA\_QUAL$>$0, and LAT\_CONFIG=1.
We further analyzed the data as described below.

\begin{figure}
\centering
\includegraphics[width=3.3 in]{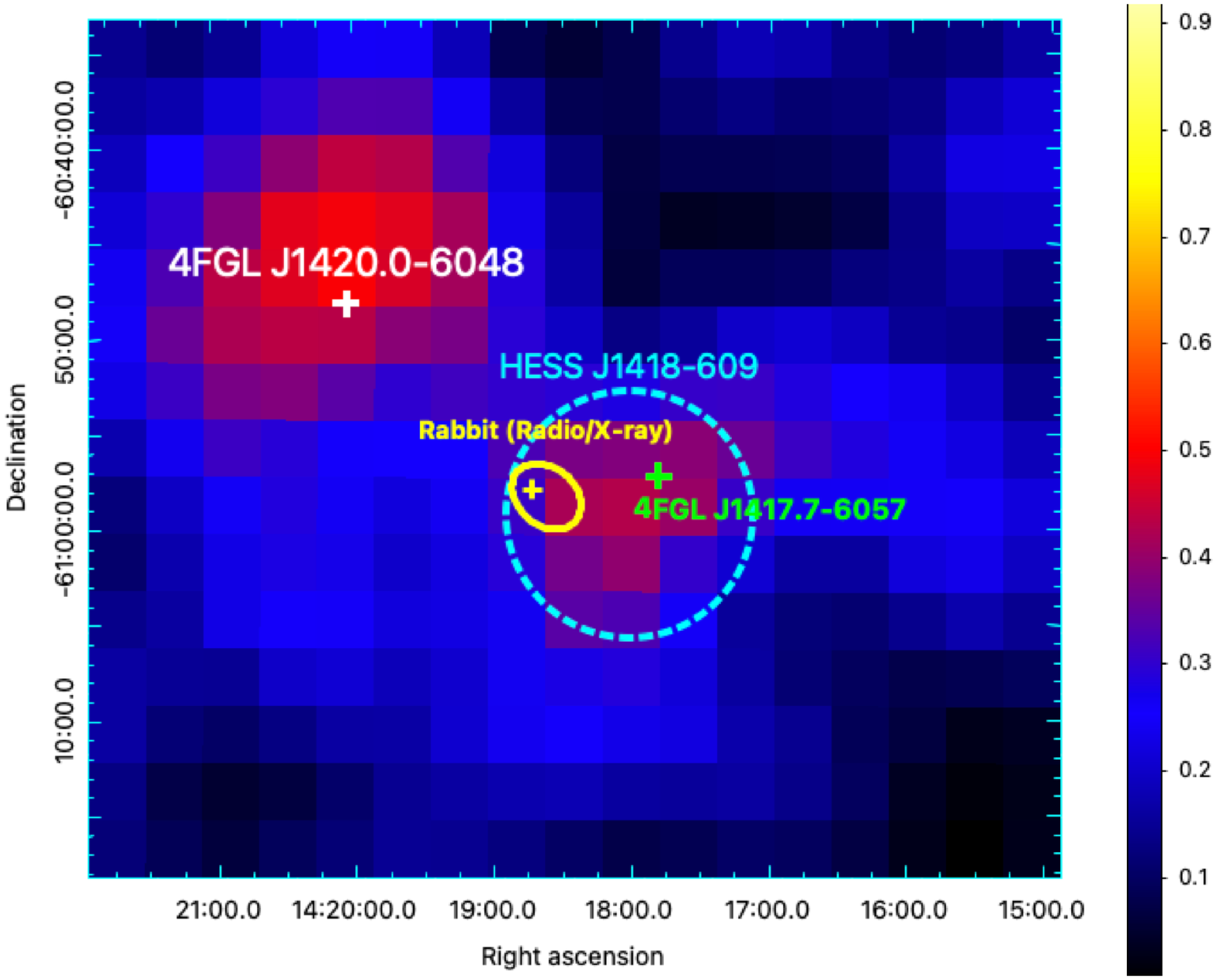}\\
\includegraphics[width=3.3 in]{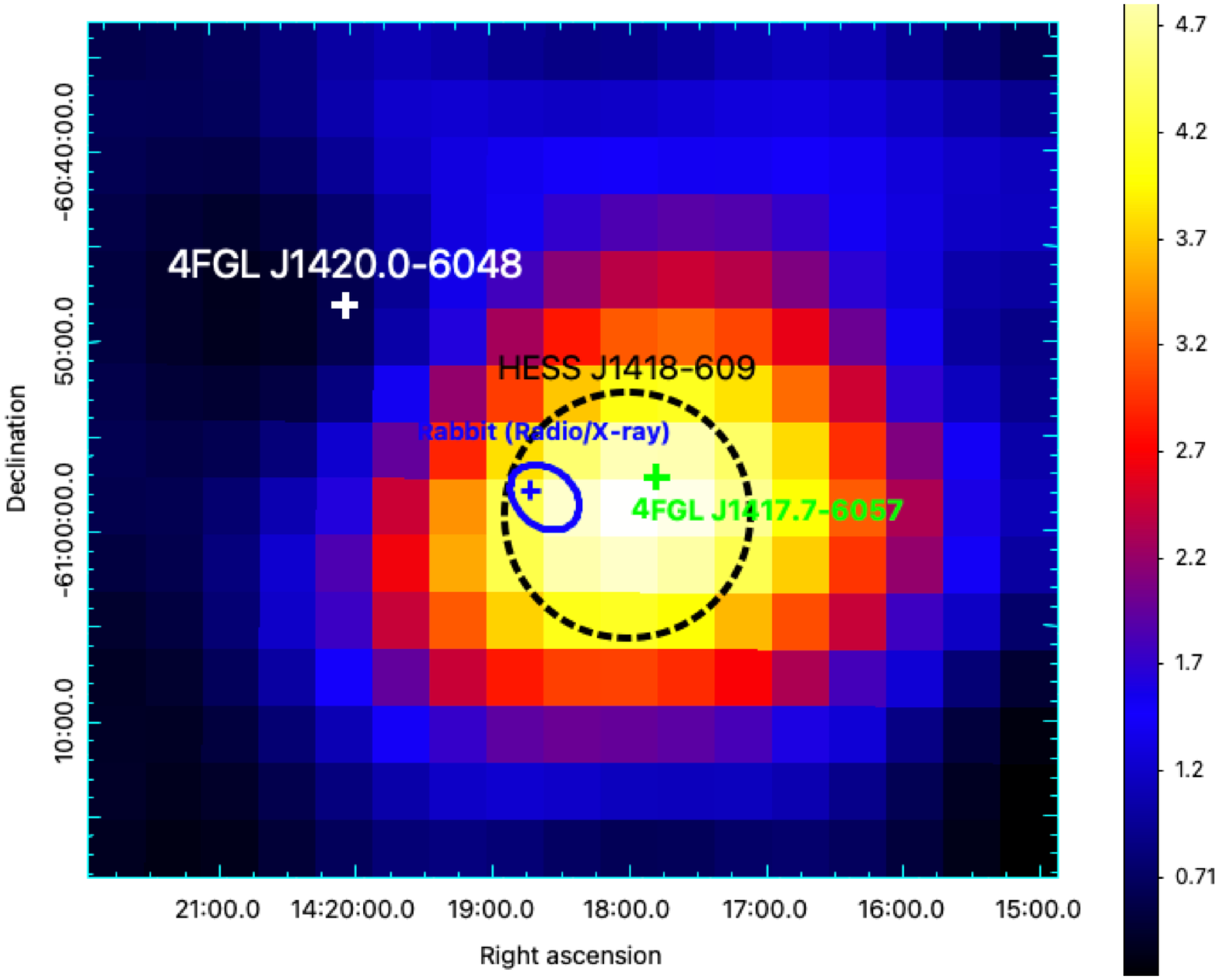}
\figcaption{A Fermi-LAT image (top) and a $\sqrt{TS}$ map (TS map; bottom)
of a $\sim 1^\circ \times 1^\circ$ region at $>$30\,GeV.
The TS map was produced by omitting the model for 4FGL~J1417.7$-$6057. The ellipses show the
Rabbit radio/X-ray PWN and the dashed circles denote the TeV source HESS~J1418$-$609
associated with Rabbit. The crosses mark a few LAT point sources: 4FGL~J1417.7$-$6057 (green),
PSR~J1420$-$6048 (white), and J1418 (cross within the ellipses).
The images are smoothed, and the scales are adjusted for legibility.
\label{fig:fig7}}
\vspace{0mm}
\end{figure}

We performed a binned likelihood analysis using $0.05^\circ$ bins
in the 100\,MeV--1\,TeV band to measure the source spectrum.
Because our data are not very different from those used for the 4FGL DR-2
catalog \citep[{\tt gll\_psc\_v27.fit};][]{fermi4fgl},
the 4FGL values are expected to be accurate.
Nevertheless, we verified the parameter values below.
We created an XML model including all the 4FGL DR-2 sources within a square
of $30^\circ\times 30^\circ$ using the parameters given in the catalog.
We started by optimizing parameters for J1417 and normalizations
for the diffuse emissions ({\tt gll\_iem\_v07} and
{\tt iso\_P8R3\_SOURCE\_V3\_v1}).\footnote{https://fermi.gsfc.nasa.gov/ssc/data/access/lat/Background\\Models.html}
We then gradually increased the number of sources to fit until no excess or deficit was identifiable in the residual plot. 
Our optimized parameters were fully consistent with the 4FGL DR-2 values.

We next inspected a LAT image in the high-energy band ($>$30\,GeV) where the 4FGL DR-2 catalog
found significant emission (i.e., J1417). We produced a count map in the $>$30\,GeV band
and found that there were excess counts in the $R\sim 0.1^\circ$ region of the H.E.S.S
counterpart (HESS~J1418$-$609; Fig.~\ref{fig:fig7} top). 
To verify the excess, we generated a TS map of the region. We removed 
J1417 from our optimized model and ran the {\tt tsmap} tool of Fermipy to generate
a $>$30\,GeV $\sqrt{TS}$ map (TS map) which is displayed in Figure~\ref{fig:fig7} bottom.
The TS map showed significant emission in the same region as that of the count map.
Most of the excess in the TS map was resolved when J1417 was included in the model.
However, it was not possible to accurately determine the position or the extension
with the data, due to poor photon statistics and/or broad LAT PSF
(i.e., 68\% containment radius of
0.1$^\circ$ at $>$30\,GeV).\footnote{https://www.slac.stanford.edu/exp/glast/groups/canda/lat\_P\\erformance.htm}

\section{SED modeling}
\label{sec:sec4}
\subsection{Construction of a broadband SED and radial profiles of X-ray properties}
\label{sec:sec4_1}
	We constructed a broadband SED of the Rabbit PWN by adding the published radio and VHE
measurements to our X-ray and LAT data. The radio flux densities measured for ``Rabbit'' with
ATCA observations were taken from \citet{Roberts1999}, and we used the VHE results for
HESS~J1418$-$609 reported by \citet{hessrabbit2006}.
Note that the X-ray and VHE fluxes are measured from different regions.
We used a $R=3'$ region for the X-ray SED, but the VHE SED was measured
from a larger region (e.g., H.E.S.S. size). In our modeling, we integrate the model emission
over the sizes appropriate for the X-ray and VHE SEDs and compare it with the
observed data. The radio flux-density measurements may not
be accurate due to possible contamination from the Kookaburra complex and incomplete UV coverage.
For SED modeling, we took the radio data at face value, although model conclusions
are not strongly influenced by those constraints.

\subsection{Multizone emission model for the PWN}
\label{sec:sec4_2}
\begin{figure*}
\centering
\includegraphics[width=6.3 in]{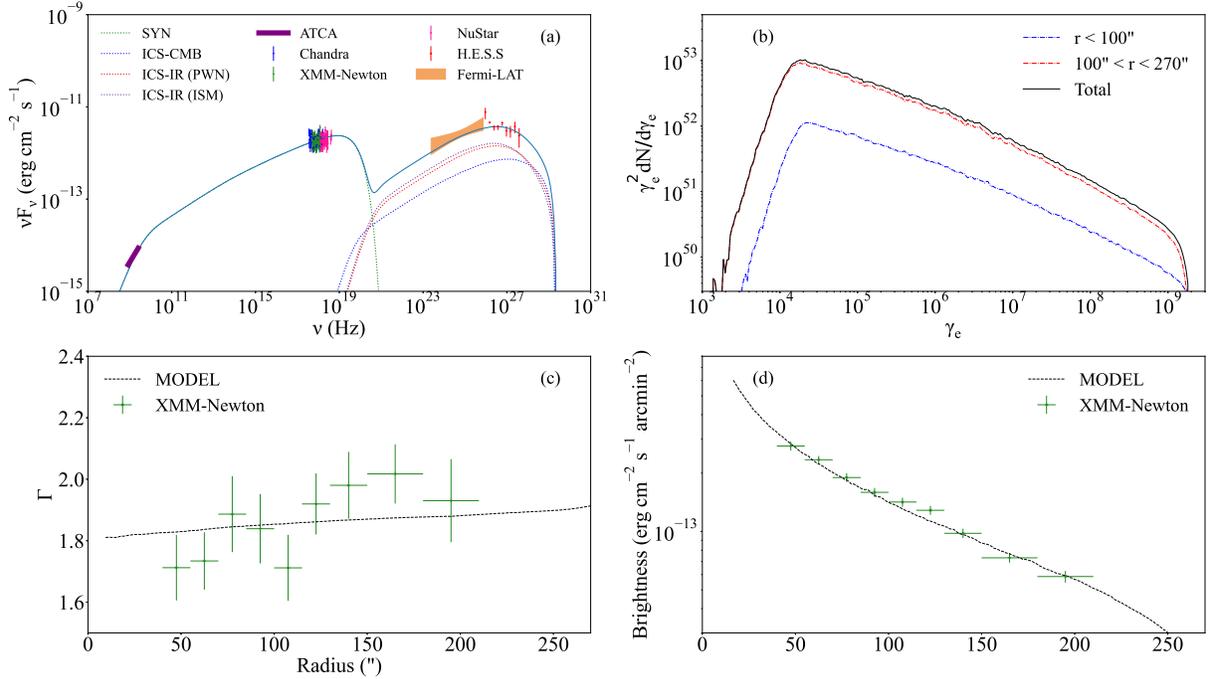}
\figcaption{Measurements and models for broadband SED and radial profiles of X-ray photon index and brightness.
({\it a}) a broadband SED (data points) and emission models.
The green dotted line is the synchrotron emission model, and
the blue, red, and purple dotted lines are models for ICS emission from CMB and IR background in the PWN
and in the ISM regions, respectively. The cyan solid line shows the summed model.
({\it b}) electron distributions in the inner ($r<100''$; blue), the outer ($100''<r<270''$; red),
and the whole region (black),
({\it c}--{\it d}) radial profiles of the photon index (c) and brightness (d) in the X-ray band.
\label{fig:fig8}
}
\vspace{0mm}
\end{figure*}

	We describe a multizone SED model \citep[e.g.,][]{KimS2020} 
that we use to infer the particle energies and the flow properties in the Rabbit PWN
by simultaneously fitting the radial profiles and the broadband SED (Fig.~\ref{fig:fig8}).
The observed morphology of the PWN is asymmetric,
and thus our model with the assumption of spherical or conical flow
is only approximate (see Section~\ref{sec:sec5_2}).

	In the model, electrons characterized by a power-law energy distribution 
\begin{equation}
\label{eq:particle}
\frac{dN}{d\gamma_e dt}=N_0 \gamma_e^{-p_1},\ \ \ \gamma_{e,\rm min} < \gamma_e < \gamma_{e,\rm max},
\end{equation}
where
$\gamma_e$ is the electron Lorentz factor, are injected into the termination shock at $R_{\rm TS}$ 
and flow in the PWN. The particle injection power
\begin{equation}
\label{eq:particleE}
\dot E_{e,\rm inj}=\int_{\gamma_{e,\rm min}}^{\gamma_{e,\rm max}} \gamma_e m_e c^2 \frac{dN}{d\gamma_e dt} d\gamma_e,
\end{equation}
where $m_e$ is mass of an electron and $c$ is the speed of light,
is a fraction of the pulsar's luminosity ($\dot E_{e,\rm inj}=\eta L(t)$) which is assumed to decrease with time
following 
\begin{equation}
\label{eq:spindown}
L(t)=L_0\left (1+\frac{t}{\tau_0}\right )^{-\frac{n+1}{n-1}},
\end{equation}
where $\tau_0=2\tau_c/(n-1)-t_{\rm age}$ and
$n$ is the braking index \citep[assumed to be 3;][]{Gaensler2006,Gelfand2009}.
The properties within the radio/X-ray PWN are prescribed as power laws: 
\begin{equation}
\label{eq:magnetic}
B(r)=B_0 \left ( \frac{r}{R_{\rm TS}}\right )^{\alpha_B}
\end{equation}
for the magnetic field,
\begin{equation}
\label{eq:flow}
V_{\rm flow}(r)=V_0 \left (\frac{r}{R_{\rm TS}}\right )^{\alpha_V}
\end{equation}
for the bulk flow speed, and
\begin{equation}
\label{eq:diffusion}
D=D_0 \left (\frac{B}{100\mu G}\right )^{-1} \left (\frac{\gamma_e}{10^9}\right )^{\frac{1}{3}}
\end{equation}
for the diffusion.
We assumed $\alpha_B + \alpha_V=-1$ which is valid for spherical (or conical)
flow and transverse $B$ with magnetic flux conservation.
Note that this relation could be different for other
flow geometries or $B$ configurations \citep[][]{r09}.
In our model, $R_{\rm TS}$ and $R_{\rm PWN}$ are constant in time,
and thus $B$ and $V_{\rm flow}$ are also constant in time. However, particles
with different ages experience different $B$, and $V_{\rm flow}$ in our model as
the particles are at different radial positions (see below).

Larger VHE emission regions compared to their radio/X-ray emission zones
observed in some PWNe (as in Rabbit) indicate that particles outside the compact X-ray PWN
can produce VHE emission via ICS. To account for this,
we assumed that the flow bulk motion ($V_{\rm flow}$)
carries particles and $B$ only out to the boundary of the X-ray PWN
(at $r=R_{\rm PWN}$): i.e., we set $V_{\rm flow}(r>R_{\rm PWN})=0$ and
assign a small value for $B(r>R_{\rm PWN})\equiv B_{\rm ext}$ (e.g., 1.5 $\mu$G).
As a result, such $B$ and $V_{\rm flow}$ values cause a discontinuity at the outer boundary of the X-ray PWN
(e.g., $B(R_{\rm PWN})=5.7\mu$G  and $V_{\rm flow}(R_{\rm PWN})=530\rm \ km\ s^{-1}$
for the parameters in Table~\ref{ta:ta2}), and so we connected the parameter values
between the inside and outside regions, using a rapidly decreasing function.
We verified that the exact functional form (e.g., step,
logistic or exponential function) does not alter the resulting emission significantly
as long as $R_{\rm PWN}$ covers the emission zones that we model.

$R_{\rm PWN}$ is prescribed to represent the PWN region with intense X-ray emission.
Without a sharp edge in the X-ray image, it is very difficult to determine
the $R_{\rm PWN}$ value observationally,
but at the same time, the value does not have a large influence on the model as long as
the `bright' X-ray emission zone is included within $R_{\rm PWN}$ (see Section~\ref{sec:sec4_3}).
Note that the sudden drop of $B$ is necessary (and sufficient) to match
the size of bright X-ray emission of PWNe with a sharp boundary (e.g., Crab)
by suppressing
synchrotron emission at $r>R_{\rm PWN}$ but
the assumption of $V_{\rm flow}(r>R_{\rm PWN})=0$ does not have a large impact on the emission in our model.
The particles can propagate farther out (into the ISM) via diffusion and produce ICS emission.

	The particle flow was computed using Monte-Carlo simulations \citep[e.g.,][]{tc12}. At each time step ($dt$)
particles moved radially outward by $V_{\rm flow}(r)dt$, randomly diffused by $\sqrt{2Ddt}$ in each direction,
and cooled via adiabatic expansion, synchrotron radiation in randomly oriented $B$, and ICS radiation off of
the CMB ($T_{\rm CMB}=2.7$\,K and $u_{\rm CMB}=0.26\rm \ eV\ cm^{-3}$)
and ambient IR photons ($T_{\rm IR}$ and $u_{\rm IR}$). 
This process was repeated over the assumed age of the PWN ($t_{\rm age}$).
We assumed reflecting and transmitting boundary conditions at
the inner ($r=R_{\rm TS}$) and the outer boundaries ($r=R_{\rm PWN}$)
of the X-ray PWN, respectively.
The synchrotron and ICS emissions of the isotropic particles were computed using
the formulae given in \citet[][]{fdb08} at each time and position.
We then calculated integrated SED and radial profiles by projecting the emission
onto the tangent plane of the observer.

	In the model, there are many free parameters: e.g., $R_{\rm TS}$, $R_{\rm PWN}$,
$p_1$, $\gamma_{e,\rm min}$, $\gamma_{e,\rm max}$,
$t_{\rm age}$, $B_0$, $\alpha_{\rm B}$, $V_0$, $\alpha_V$ ($=-1-\alpha_B$), $D_0$, $T_{\rm IR}$ and $u_{\rm IR}$. 
Some of the parameters can be tightly constrained by the observation data (e.g., $R_{\rm PWN}$ if a sharp
X-ray edge is detected).
The shapes of the synchrotron and ICS SEDs are primarily controlled by 
the injected particle distribution ($p_1$, $\gamma_{e,\rm min}$, and $\gamma_{e,\rm max}$; eq.~\ref{eq:particle}),
$B$ and $V_{\rm flow}$ (eqs.~\ref{eq:magnetic} and \ref{eq:flow}).
The synchrotron cooling is dominant for the highest-energy electrons (near $\gamma_{e,\rm max}$)
and thus determines the synchrotron SED shape in the hard X-ray band,
whereas adiabatic cooling is dominant for low-energy electrons
and therefore is relevant to the low-energy SED.
The parameters for the ambient IR field ($T_{\rm IR}$ and $u_{\rm IR}$) were 
adjusted to match the amplitude of the TeV SED, and $V_0$ ($\alpha_B$ and $\alpha_V$),
and $D_0$ are adjusted to match the radial profiles of the X-ray brightness and photon index.

For the parameter optimization, we iteratively adjusted the model parameters until
a good match between the model and the measurements was achieved (visual inspection).
We then carried out pair scans of important parameters to refine our
parameter estimations, employing the $\chi^2$ statistic (Figs.~\ref{fig:fig8} and \ref{fig:fig9}). 
Because it is uncertain whether or not
all of the radio and TeV emissions are associated with the X-ray PWN, and
the radio and TeV data can be matched relatively easily by simply adjusting $\gamma_{e,\rm min}$
and $u_{\rm IR}$ without altering the other important parameters
(e.g., $\gamma_{e,\rm max}$ and $B_0$), we used only the X-ray measurements for the fit.
In principle, a change of $\gamma_{e,\rm min}$
results in a slight change in the total particle energy ($\eta$).
Then to keep $\eta$ constant while matching the X-ray data,
we need to adjust $B_0$, but the required change of $B_0$ is
small for a modest change of $\gamma_{e,\rm min}$. We note that a sharp cutoff in the electron energy distribution below some $\gamma_e$ is not expected in most theories of particle acceleration at shocks.  More detailed radio observations of K3 at 1 GHz and below could be of significant value in improving our understanding of PWN radio emission in general.
Further note that the radial profiles measured by
XMM-Newton, Chandra, and NuSTAR (Fig.~\ref{fig:fig6} right) show large scatter due to
the cross-normalization issues (Section~\ref{sec:sec2_5_2}), and thus matching them all with a model is not possible. We, therefore, used the XMM-Newton measurements of the radial profiles for the pair scans. Since the broadband X-ray SED, especially the NuSTAR measurement, is crucial for the estimation of $\gamma_{e,\rm max}$, we used all the spectral measurements in Fig.~\ref{fig:fig5} after normalizing the flux to the XMM-Newton flux of the $40''<R<3'$ region.

\subsection{Application of the model to the Rabbit PWN}
\label{sec:sec4_3}
In a previous study, a one-zone time-dependent model \citep[][]{Zhu2018} was used to explain
broadband SEDs of several PWNe (including the Rabbit PWN). We used their model parameters as a guide to our model input. 
Note, however, that the one-zone modeling did not account for the spatial variation of the X-ray spectra
within the PWNe, and hence the parameters needed to be modified in our multi-zone model.
The true age of the Rabbit PWN is unknown, but a correlation between the
X-ray-to-gamma-ray luminosity ratio and age of a large ensemble of PWNe \citep[][]{Kargaltsev2013,Zhu2018}
suggests several kyr for Rabbit's true age.
We assumed an age of 7000\,yr, $R_{\rm TS}=0.1$\,pc, and
$R_{\rm PWN}=$4.6\,pc (i.e.,
X-ray emission region of $\approx 4.5'$ for an assumed $d$=3.5\,kpc).
Multi-zone emission models computed with $10^5$ temporal, 1000 spatial,
and $10^4$ energy bins are plotted in Figure~\ref{fig:fig8},
and the model parameters are presented in Table~\ref{ta:ta2}.

	The spectral shape of the synchrotron emission by uncooled electrons (Fig.~\ref{fig:fig8} a),
relevant to $p_1$, is not well measured,
but the LAT SED suggests a hard power law for the electron distribution and $p_1=$2.27
adequately explains both the X-ray and VHE SEDs (Fig.~\ref{fig:fig8}).
The similar radio and X-ray sizes of the Rabbit PWN already suggest
that $B$ is weak in the source \citep[as in G21.5$-$0.9;][]{Matheson2010}.
The insignificant softening of the X-ray spectrum
(Sections~\ref{sec:sec2_4_2} and \ref{sec:sec2_5_4}) further
implies that diffusion is efficient in the source. Our best-fit $B_0$ of 12$\mu$G is similar
to that inferred from the one-zone modeling \citep[$4.4\mu \rm G$;][]{Zhu2018}.
For this $B_0$, the synchrotron cooling timescale for the X-ray emitting electrons ($\gamma_e\approx10^8$--$10^9$)
is $\le10^3$\,yrs, and hence they
cool substantially over 7000\,yr. However, since the efficient diffusion compensates for
the cooling, the particle spectral variation with a distance from the pulsar
is not large (Fig.~\ref{fig:fig8} b).
Alternatively, the insignificant spectral softening may be explained by a pure
advection model which predicts nearly constant $\Gamma$ out to a certain radius and a rapidly increasing trend at large radii in a 1D case \citep[e.g.,][]{Reynolds2003} if the cooling break
is above the observed X-ray band in the inner regions. This requires low $B$ ($< 3\mu\rm G$) for the assumed age of 7000\,yr of Rabbit, and then
$N_0$ and thus $\eta$ should increase by more than an order of
magnitude to fit the X-ray SED. We do not consider this
case since it is very difficult to substantially increase the
injected particle energy for the given $L_{\rm SD}$ of J1418.

Adiabatic cooling of the low-energy particles accounts for the hard radio
SED without requiring an intrinsic spectral break in the particle distribution.
Radio fluxes are controlled primarily by $\gamma_{e,\rm min}$, and
the radio SED slope is related to particle cooling (e.g., $B$ and $V_{\rm flow}$).
Because of uncertainties in the radio measurements,
we matched them with our model only by visual inspection.
As noted above, the radial profile of the X-ray photon index (Fig.~\ref{fig:fig8} c)
suggests fast diffusion in the PWN \citep[see also][]{tc12,VanEtten2011}.
We find that $D_0\approx 10^{27}\rm \ cm^2\ s^{-1}$
can adequately explain the measured $\Gamma$ profile.
The brightness profile (Fig.~\ref{fig:fig8} d) implies that the radial decrease of $B$ is small
($\alpha_B=-0.2$).

	Since the X-ray PWN does not have a sharp boundary, our choice of the $R_{\rm PWN}$ value
is rather arbitrary although the bright part of the PWN is included well within the radius.
As noted above, this is related to the sudden drop of $B$ and $V_{\rm flow}$ in our model.
The former has some influence on the results. For larger $R_{\rm PWN}$, $B$ and brightness in outer
zones are higher, making the brightness profile flatter; this can be accommodated by
our model with changes of the other parameters (e.g., $B_0$, and $\alpha_B$).
Since $B$ and brightness in outer zones are already low, the required changes of the parameters
to preserve a data-vs-model match are not large. For an increase of $R_{\rm PWN}$ by a factor of $\sim$1.5,
the other parameters need to be changed $\le$20\%. Note that
the parameter dependence on $R_{\rm PWN}$ is not linear
because emission in far outer regions is much weaker.

	To match the TeV SED, we adjusted the temperature
\citep[e.g., $T_{\rm dust}$=10--30\,K;][]{Zhu2014} and density of the IR field,
as has been often done in SED modelings \citep[e.g.,][]{tcm13,Zhu2018}.
The TeV emission of Rabbit is detected over a more extended region than the lower energy counterparts \citep[][]{HESS2018},
indicating that particles that escaped from the X-ray PWN (e.g., $R_{\rm PWN}=4.5'$) into ISM upscatter
IR photons there.  The computed TeV SEDs within the X-ray PWN (red) and in the ISM (purple)
are separately presented in Figure~\ref{fig:fig8} (a).
Note that we assumed the IR field is spatially homogeneous. However, bright mid-IR emission at 8--20$\mu m$ was
detected $\sim$2--3$'$ west from the pulsar in the WISE and Glimpse images.
If this mid-IR source is associated with or at the same $d$ as the Rabbit PWN,
the ICS emission would have a SED bump at $\sim$100\,GeV which we do not see in the VHE data.
Moreover, the VHE emission would be stronger at the location of the mid-IR source
because its emission is intense. The LAT and TeV
counterparts of Rabbit are far away from the mid-IR source. 
These suggest that the mid-IR source may not be associated with the PWN.
It is also possible that some of the VHE emission is produced by sources other than the PWN,
e.g., a putative supernova remnant (SNR) shell. In our model, this would imply
a lower external IR density ($u_{\rm IR}$) or alternatively higher $B$ and a smaller number of particles
for fixed $u_{\rm IR}$.

\begin{table}[t]
\vspace{-0.0in}
\begin{center}
\caption{Parameters for the multizone SED model}
\label{ta:ta2}
\vspace{-0.05in}
\scriptsize{
\begin{tabular}{lcc} \hline\hline
Parameter  & Symbol  & Value   \\ \hline
Spin-down power (today) & $L_{\rm SD}$              & $5\times 10^{36}\rm \ erg\ s^{-1}$     \\
Characteristic age of the pulsar & $\tau_{\rm c}$   & 10400\,yr        \\
Age of the PWN        & $t_{\rm age}$              & 7000\,yr        \\
Size of the PWN       & $R_{\rm pwn}$  & 4.6\,pc        \\
Radius of termination shock & $R_{\rm TS}$  & 0.1\,pc        \\
Distance to the PWN  & $d$  & 3.5\,kpc        \\ \hline
Index for the particle distribution   & $p_1$            & 2.27          \\
Minimum Lorentz factor  & $\gamma_{e,\rm min}$  & $10^{4.36}$        \\
Maximum Lorentz factor  & $\gamma_{e,\rm max}$ & $10^{9.3}$        \\
Magnetic field        & $B_0$      & 12.3$\mu$G       \\
Magnetic field at $r>R_{\rm PWN}$  & $B_{\rm ext}$      & 1.5$\mu$G        \\
Magnetic index        & $\alpha_B$   & $-$0.2 \\
Flow speed            & $V_0$        & 0.038$c$ \\
Speed index           & $\alpha_V$    & $-0.8$ \\
Diffusion coefficient & $D_0$    & $1.1\times10^{27}\rm \ cm^2 \ s^{-1}$ \\
Energy fraction injected into particles   &  $\eta$      &  0.88        \\
Energy fraction injected into $B$ field        & $\eta_{B}$    & 0.0036      \\ \hline
Temperature of IR seeds  & $T_{\rm IR}$  & 20\,K        \\
Energy density of IR seeds  & $u_{\rm IR}$  & 2.2$\rm \ eV \ cm^{-3}$   \\ 
CMB temperature  & $T_{\rm CMB}$  & 2.7\,K   \\
CMB energy density & $u_{\rm CMB}$  & 0.26$\rm \ eV \ cm^{-3}$   \\ \hline
\end{tabular}}
\end{center}
\vspace{-0.5 mm}
\end{table}

\subsection{Model Parameter Covariance}
\label{sec:sec4_4}

\begin{figure*}
\centering
\includegraphics[width=7 in]{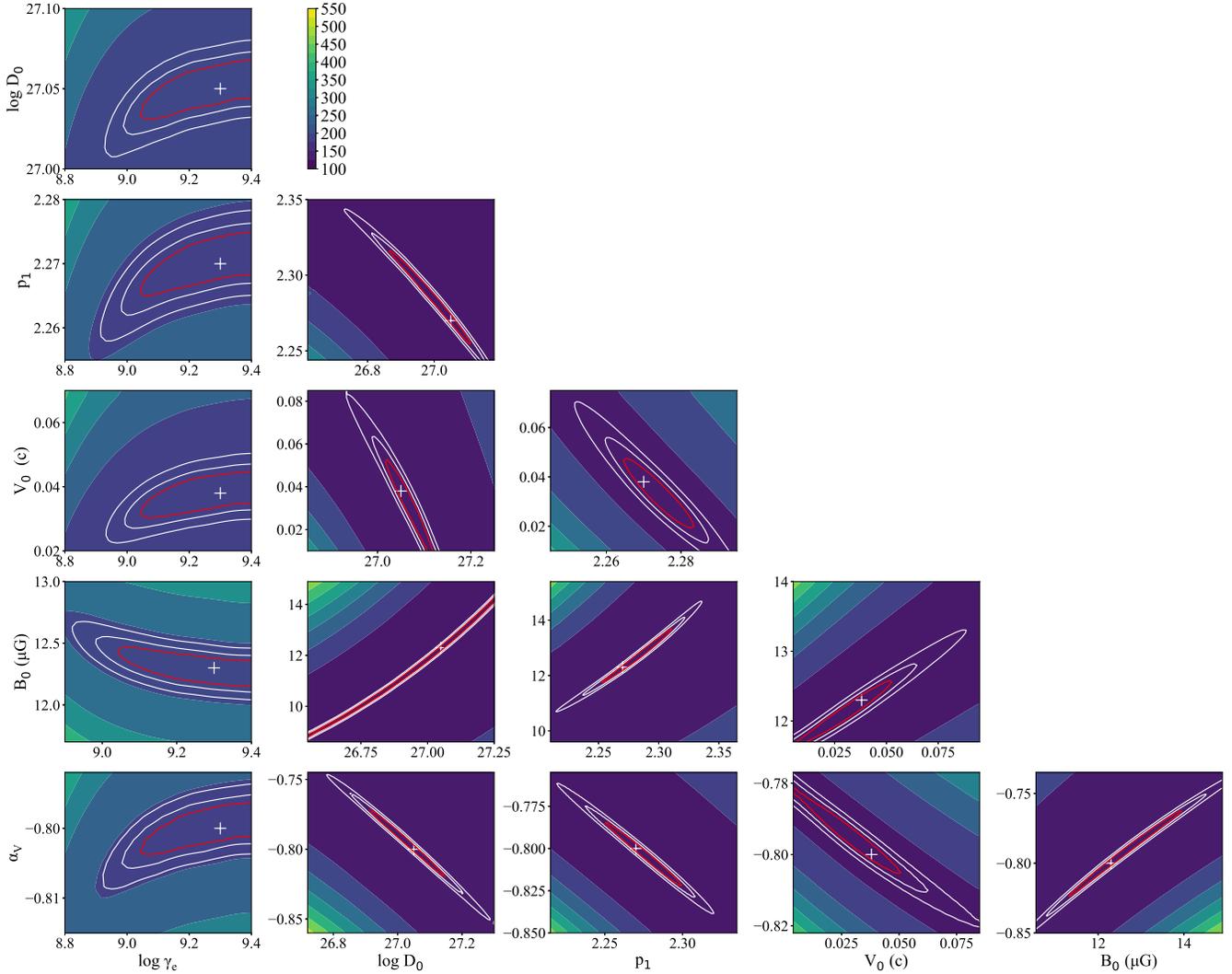}
\figcaption{$\chi^2$ contours obtained by pair scans. 68\%, 90\%, and 99\% contours are displayed.
The designated pair of the model parameters are simultaneously scanned
while holding the other parameters fixed at their best-fit
values (Table~\ref{ta:ta2}), and $\chi^2$ is computed with the X-ray SED
and radial profiles of $\Gamma$ and brightness. The crosses mark
the parameter values reported in Table~\ref{ta:ta2}.
\label{fig:fig9}
}
\vspace{0mm}
\end{figure*}

	Optimization of the model parameters and inspection of covariance between them
require multi-dimensional parameter scans, which are computationally unfeasible.
Hence, we instead carried out pair scans for several important parameters.
Note that the other parameters are frozen because they do not significantly influence the
X-ray emission (e.g., $R_{\rm TS}$, $\gamma_{e,\rm min}$, $T_{\rm IR}$, and $u_{\rm IR}$)
or could be determined by images (e.g., $R_{\rm TS}$ and $R_{\rm PWN}$) in principle.
For a pair of parameters, we varied their values around those determined by visual inspection
and computed $\chi^2$ by fitting the X-ray data. The results of the pair scans are presented in Figure~\ref{fig:fig9}.
Because we did not simultaneously optimize the parameters, the reported (our optimized) parameters are off-centered in some of the panels but within the 68\% contours.

	The complex interplay among the parameters is not fully captured by the pair scans,
but they show some general covariances which could be qualitatively understood as follows.
The integrated flux is mainly determined by particle residence time and $B$.
For small $V_0$, $\alpha_V$ (large negative value), and $D_0$, the residence time is long
and thus emission within the PWN volume is large. $\alpha_V$ and $D_0$ further
control the radial profiles with
some complications due to the $\alpha_V + \alpha_B=-1$ relation and energy-dependent diffusion;
in general, smaller values produce more rapidly falling brightness and $\Gamma$ profiles.
The X-ray spectral shape is affected by $p_1$, $B_0$ and $\gamma_{e,\rm max}$ as
the synchrotron emission frequency ($\nu_{\rm syn}$) is proportional to $B_0 \gamma_e^2$.

$\gamma_{e,\rm max}$ is weakly correlated with $D_0$, $p_1$, $V_0$ and $\alpha_V$,
and anti-correlated with $B_0$. Larger $\gamma_{e,\rm max}$ does not degrade the SED fit
because there are no high-energy data ($>$20\,keV). In this case, however, the $\Gamma$ profile
becomes flatter by diffusion/advection of the higher-energy particles to outer regions.
To keep the fit quality, these higher-energy particles need to be pushed out to low-$B$ regions by more rapid propagation (larger $D_0$, $V_0$ or $\alpha_V$), or alternatively, 
the injected particle spectrum should be softer (larger $p_1$).
The $B_0$-$\gamma_{e,\rm max}$ anti-correlation is obvious since $\nu_{\rm SY}\propto B \gamma_e^2$.
$D_0$ is correlated with $B_0$ but anti-correlated with $\alpha_V$, $V_0$ and $p_1$.
The decrease of the total PWN emission caused by shorter residence time (larger $D_0$)
is balanced by larger $B_0$ and/or smaller $\alpha_V$ and $V_0$. An increased loss of high-energy particles in outer regions,
due to stronger diffusion, makes the X-ray spectrum softer, which is compensated by smaller $p_1$ (harder injection spectrum).
The $p_1$-$V_0$, $p_1$-$B_0$, and $p_1$-$\alpha_V$ correlations are seen because larger $p_1$
means less X-ray emitting particles for given $\eta$;
to explain the observed X-ray flux, higher $B_0$ or longer residence time is necessary.
The $V_0$-$B_0$, $V_0$-$\alpha_V$ and $B_0$-$\alpha_V$ correlations are also related to the total emission
within the PWN.

\section{Discussion and Conclusions}
\label{sec:sec5}
	We determined broadband X-ray spectra of the Rabbit PWN using archival
Chandra and XMM-Newton data, and a new NuSTAR observation. NuSTAR's high temporal resolution
allowed us to detect 110-ms X-ray pulsations of J1418 with high significance,
characterize the pulse profile in the hard X-ray band and clearly distinguish between 
the on- and off-pulse emissions. 
By jointly analyzing Chandra, XMM-Newton, and NuSTAR's off-pulse data,
we found that the X-ray spectrum of the PWN is well described by a power-law model with 
$\Gamma\approx2$.
We then applied a multizone emission model to investigate flow properties in the PWN
and found that the electrons are accelerated to very high energies ($\gapp$500\,TeV)
in the Rabbit PWN.

\subsection{Observed X-ray and VHE properties of the Rabbit PWN}
\label{sec:sec5_1}
	While the detection of X-ray pulsations of J1418 was claimed
in a previous XMM-Newton study \citep[][]{Kim2020}, the
significance was not very high, and the pulse profile was not well characterized.
The NuSTAR confirmation of the pulsations from J1418 firmly 
established its association with the Rabbit PWN. 
Furthermore, we found that the X-ray pulse profile of J1418 exhibits
a sharp peak and a broad bump with a phase separation of 0.5.
The pulsar's gamma-ray light curve also shows two peaks with the same phase separation.
A comparison between the gamma-ray and X-ray pulse profiles can lead to determining the spin orientation
of the pulsar \citep[e.g.,][]{Wang2013}; the previous XMM-Newton study
found that the sharp X-ray peak in the profile phase-aligns well with a GeV peak.
A further investigation with our NuSTAR measurement requires an accurate LAT timing solution
that covers the NuSTAR observation epoch. 

	The accurate characterization of the X-ray pulse profile with the NuSTAR data allowed
an investigation of the hard X-ray emission properties of the PWN.
Joint spectral analyses of the XMM-Newton, Chandra, and NuSTAR data demonstrated that
the spatially integrated X-ray spectrum of the PWN is well described by
an absorbed power law with $N_{\rm H}=(2.78\pm 0.12)\times 10^{22}\rm \ cm^{-2}$ and
a photon index $\Gamma=2.02\pm0.05$. These are consistent with those measured by
Suzaku \citep[$\Gamma=2.00\pm0.06$ for $N_{\rm H}=2.7\times 10^{22}\rm \ cm^{-2}$;][]{Kishishita2012}.
Note, however, that they did not report the abundance and cross-sections used for their
Galactic absorption model, and hence we assumed that they used the {\tt angr} abundance
and {\tt vern} cross section (the default in XSPEC).
In the Suzaku data, \citet{Kishishita2012} found a significant
spectral softening, with $\Gamma$ gradually growing from 1.77 in the inner region ($R<0.8'$)
to 2.12 in the outer region ($R=1.7'$--$3'$), which is not consistent with the results of
our imaging analysis (Section~\ref{sec:sec2_4})
as well as spatially-resolved spectral analysis (Section~\ref{sec:sec2_5_4}).
Most likely, the contamination by the pulsar (and possibly X1) was not adequately
removed in the Suzaku data analysis, and the hard pulsar emission may have
contaminated  their PWN spectra and caused the spectral softening (see Section~\ref{sec:sec2_5_4}).

We found that the high-energy ($>$30\,GeV) emission in the Rabbit regions appears
to overlap well with the $R=0.11^\circ$ H.E.S.S source (Fig.~\ref{fig:fig7})
and that the $>$30\,GeV SED  connects well  to the VHE one (Fig.~\ref{fig:fig8} a).
This verifies that the LAT and the H.E.S.S sources are indeed associated with each other.
Then, the LAT source should be also extended since the ICS mechanism produces gamma-ray emission in
the LAT and the VHE bands, although the LAT extension could not
be clearly constrained with the current data due to the paucity of counts and the broad LAT PSF.

\subsection{Modeling of the broadband emission properties}
\label{sec:sec5_2}
It was suggested that the one-sided morphology and the offset TeV emission
of Rabbit might be caused by a reverse shock interaction which diverts the
particles in the direction opposite to the interaction site
with respect to the pulsar \citep[e.g.,][]{hessrabbit2006}.
The interaction would complicate the flow geometry and
thus require MHD simulations that incorporate detailed physics of
the reverse shock interaction and the subsequent flow, which
is beyond the scope of our phenomenological emission model.
For non-spherical (or non-conical) flow, the adiabatic cooling and the $\alpha_{\rm B}$-$\alpha_{\rm V}$
relation may differ \citep[e.g.,][]{r09}, and thus different values for
$\gamma_{e,\rm min}$, $\alpha_B$, and $\alpha_V$ may be inferred.
In addition, particle propagation and PWN properties in the direction toward the presumed reverse shock
(i.e., the north-east direction for Rabbit) may be different from those in the X-ray
PWN (i.e., the southwest tail of the Rabbit nebula). 
Our model assumes that the same electron population
produces both the X-ray synchrotron nebula and TeV emission, after accounting for energy
losses and diffusion. In reality, it could occur that, for instance,  
the relative motion between the pulsar and its surroundings sweeps the particle
flow toward the southwest, elongating the PWN in that direction as observed \citep[e.g.,][]{Kolb2017, slane2018},
and subsequently the TeV emission further downstream.
Moreover, some of the observed emissions may be contaminated by sources (e.g., a putative SNR) other than the Rabbit PWN.
Since our simple spherical model does not account for these complex phenomena,
the reported values of the parameters need to be taken with caution.
Adequate treatments of the aforementioned complexities await further theoretical studies which we defer to future work.

\citet{Zhu2018} modeled a Rabbit SED using a one-zone time-dependent scenario,
but this one-zone model did not account for the radial profiles of the X-ray brightness and photon index.
Our multi-zone emission model with a single power-law electron spectrum
and spatially varying PWN properties reproduced the measured
SED and radial properties of the photon index and brightness well.
Although a unique set of model parameters may not be derived
from the current observational data alone due to covariance among the parameters,
there are a few interesting parameters that we could infer from the modeling.

	The large extension of the VHE emission of Rabbit 
compared to the radio/X-ray PWN (Fig.~\ref{fig:fig1} b)
can be attributed to rapid diffusion \citep[e.g.,][]{VanEtten2011};
particles that escaped from a compact X-ray PWN can give rise to TeV emission
by ICS of ambient IR photons in a much larger region.
This scenario seems plausible for the Rabbit PWN as its TeV emission lies in the direction of the X-ray
tail (along the particle outflow). Intriguingly, the diffusion length scale of
$R_{\rm diff}\sim2\sqrt{D t_{\rm age}}\approx 4\times 10^{19}\rm cm$
estimated for the VHE emitting $\gamma_e\approx 10^7$ electrons
corresponds to 0.23$^\circ$ for the assumed distance of 3.5\,kpc,
which is in accord with the extension of the TeV source from J1418 (Fig.~\ref{fig:fig1} b).

\begin{figure*}
\centering
\includegraphics[width=6.5 in]{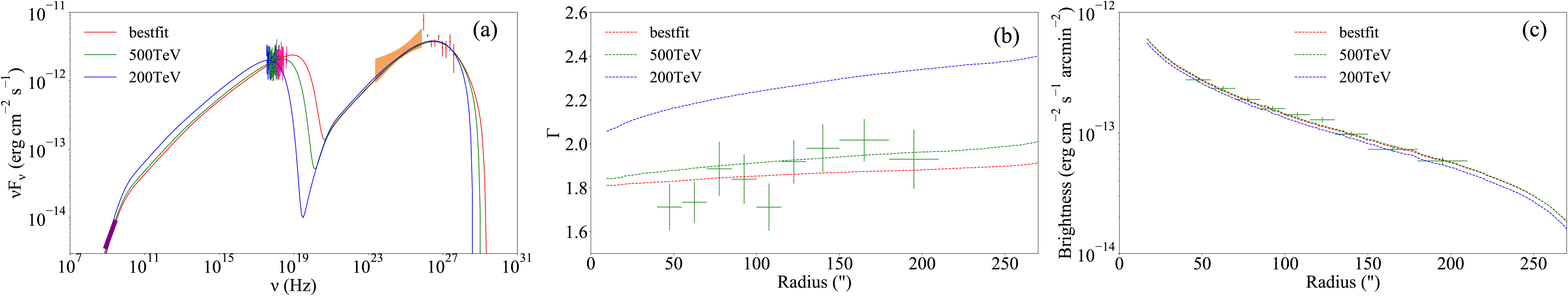}\\
\includegraphics[width=6.5 in]{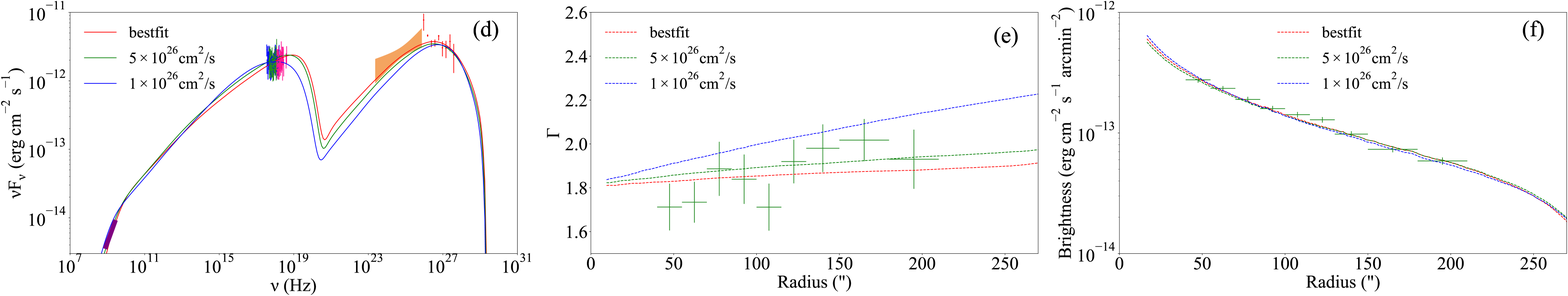}
\figcaption{Optimized models for different $\gamma_{e,\rm max}$ (a--c) and $D_0$ (d--f):
broadband SEDs (a and d), and radial profiles of $\Gamma$ (b and e) and brightness (c and f).
$\gamma_{e,\rm max}$ or $D_0$ was held fixed at the designated values
, and the other parameters were optimized by visual inspection.
\label{fig:fig10}
}
\end{figure*}

Our measurements and multi-zone modeling provide further insights into the Rabbit PWN.
In particular, the photon index and brightness profiles provide important clues to understanding the source.
The flat X-ray photon-index profile is hard to explain without rapid diffusion
($\approx 10^{27}\rm \ cm^2\ s^{-1}$) because the synchrotron and ICS cooling is severe
even for low $B$ ($\sim10\ \mu$G). 
Note again that a pure advection model \citep[e.g.,][]{Reynolds2003}
with very low $B_0$ may also explain the flat $\Gamma$ profile,
but in this case, the required particle-injection power
would be greater than $L_{\rm SD}$ of J1418. While this may be
remedied by changes in the other parameters, $\gamma_{e,\rm max}$ is
unlikely to change substantially in such an alternative model.
The flat photon-index profile (i.e., insignificant softening)
and the hard X-ray emission of Rabbit helped to constrain the maximum energy of
electrons in the PWN to be $\ge$500\,TeV ($\gamma_{e,\rm max}\approx 10^9$).
A smaller $\gamma_{e,\rm max}$ and a larger $B$ may account for the hard X-ray
emission but then faster particle cooling will make it difficult to match 
the radial profiles of the photon index and brightness. 
On the other hand, a model with a larger $\gamma_{e,\rm max}$
and a smaller $B$ overpredicts the brightness at large distances.

To investigate the possibility of lower $\gamma_{e,\rm max}$ or $D_0$,
we held the parameter fixed at smaller values and optimized the other parameters
by visual inspection. Models with lower $\gamma_{e,\rm max}$ or $D_0$
are displayed in Figure~\ref{fig:fig10}. The lower $\gamma_{e,\rm max}$ models
predicted the synchrotron cut-off at lower energies, making the predicted X-ray
spectra softer (larger effective $\Gamma$ values) and thus fits to the X-ray SED
and $\Gamma$ profile poor. Models with smaller $D_0$ values could match the X-ray SED,
but the $\Gamma$ profile showed a significant softening that is different from the
measured flat trend when $D_0$ was significantly lower (Fig.~\ref{fig:fig10} e).
Moreover, the particle residence time is large, and thus
the ICS emission of the PWN is very strong. To match the VHE SED,
we need to reduce the IR seeds (i.e., $u_{\rm IR}\approx 0$
for $D_0=10^{26}\rm \ cm^{2}\ s^{-1}$ case),
but then the Fermi-LAT SED was underpredicted at low energies (Fig.~\ref{fig:fig10} d).

While the above study added some credence to our $\gamma_{e,\rm max}$ estimation, the model
parameter degeneracy was not fully explored in this estimation of $\gamma_{e,\rm max}$.
It can be directly inferred by measuring a spectral cutoff of the synchrotron emission \citep[e.g.,][]{a19}
which was not detected in the NuSTAR data we analyzed.
A sensitive hard X-ray and MeV observations beyond the NuSTAR band, with
near-future observatories such as FORCE, HEX-P, and
COSI \citep[][]{Nakazawa2018,Madsen2018,Tomsick2019}, 
will be needed.

\subsection{Comparison with another middle-aged PWN}
\label{sec:sec5_3}
We compare the properties of Rabbit to the archetypal Vela X PWN which
also exhibits one-sided tail emission and is powered by a middle-aged pulsar
with $\tau_c=11.3$\,kyr and $\dot E_{\rm SD}=7\times 10^{36}\rm \ erg\ s^{-1}$.
$\Gamma$ within the Vela X PWN seems unchanged in the inner regions (to $\sim50'$),
but an overall increase (to $\sim90'$) is noticeable along the tail \citep[e.g.,][]{slane2018}.
TeV emission was detected, and 
\cite{HESS2019VelaX} inferred that $B$ in the Vela X PWN
decreases from 8.6$\rm \mu G$ to 5.4$\rm \mu G$ over a distance of $\approx$3\,pc.
These $B$ values imply $\alpha_B\approx -0.24$ (Eq.~\ref{eq:magnetic}).
From the TeV size of the source,
\citet{HESS2019VelaX} inferred a diffusion coefficient of $D\ge10^{27} \rm \ cm^{2}\ s^{-1}$
for 1\,TeV electrons, which is consistent with $\le 10^{28}\rm \ cm^{2}\ s^{-1}$ at 10\,TeV
suggested by \citet{Huang2018}.

While these properties are generally similar to those we measured or inferred for Rabbit,
some differences are noticeable in detailed comparisons. In the Rabbit PWN,
X-ray spectral softening is less significant, and the inferred $B$ profile is flatter, and
the inferred maximum particle energy (i.e., $\gamma_{e,\rm max}$) is higher (by a factor of $\ge$2--3)
than in the Vela X PWN.
In addition, the Vela X PWN shows a compact region in the vicinity of the pulsar
which is connected to a diffuse PWN region by a narrow structure, whereas the Rabbit PWN
exhibits a broad and continuous tail.
We speculate that these differences are related to the evolutionary stage of the sources.
HD simulations \citep[][]{Kolb2017,slane2018} have shown that one-sided morphologies
are produced when the reverse shock (RS) disrupts the PWN; the predicted morphology
of a PWN in this stage appears similar to that of Rabbit
\cite[e.g., 7500-yr case in Fig.~11 of][]{slane2018}. At later times, the RS
sweeps the pulsar wind and creates a relic PWN; the Vela X PWN was suggested to be
in the relic-PWN stage \citep[Fig.~12 of][]{slane2018}.
Then, the differences in the X-ray spectral softening and $B$ profile
for the Rabbit and Vela X PWNe could be explained as due to larger $B$ contrast
in the more evolved Vela X PWN between the inner fresh-wind zone and the outer relic PWN.

The most significant difference between the two sources is whether or not
their SNRs (e.g., ejecta and shell emission) were detected;
emission of the host SNR of the Vela X PWN was identified \citep[e.g.,][]{slane2018},
whereas the emission signature of the Rabbit SNR has not been found yet.
With the lack of SNR emission, the formation of the tail
in Rabbit is puzzling in the PWN-SNR evolution scenarios \citep[e.g.,][]{Kolb2017,slane2018}.
Is the one-sided morphology of Rabbit produced by supersonic motion
of the pulsar as in bow-shock nebulae, and 
not by the RS interaction? Then the small characteristic age of J1418 and
the strong TeV emission from the Rabbit PWN are unusual
compared to other bow-shock nebulae \citep[e.g.,][]{Kargaltsev2017}.
A proper motion measurement of J1418 and/or detection of SNR emission around Rabbit,
with sensitive X-ray observatories \citep[e.g., Lynx or AXIS;][]{Gaskin2019,Mushotzky2019},
will be needed to address this issue, and 
dedicated radio studies might also cast light on the system,
refining the fluxes attributable to the PWN and perhaps locating the SNR shell.
Furthermore, observational and theoretical studies of
these two and other middle-aged PWNe
can provide insights into the evolution of PWNe and their interaction
with the SNR and ambient medium.

\section{Summary}
\label{sec:sec6}
We characterized the emission properties of J1418 and Rabbit, and applied a multi-zone model to the
measurements. Below we summarized our main conclusions.
\begin{itemize}
\item We found that the X-ray pulse profile of J1418 exhibits a sharp peak and a broad
bump separated by $\approx$0.5 phase.
\item We found out that the 0.5--20\,keV spectrum of the Rabbit PWN is well modeled by
a $\Gamma\approx2$ power law and does not significantly soften with
increasing distance from the pulsar.
\item Our multi-zone modeling of the broadband SED and the radial profiles of $\Gamma$
and brightness of the PWN suggests that its magnetic field is low ($\sim$10$\mu\rm G$),
and the particles are accelerated to very high energies ($\gapp$500\,TeV) and diffuse out efficiently
($D\sim 10^{27}\rm \ cm^2\ s^{-1}$).
\end{itemize}

	As noted above, the flat radial profile of the photon index requires both low $B$
and efficient diffusion. We have assumed magnetic flux conservation ($\alpha_V + \alpha_B = -1$),
but the magnetic field could either be dissipated by reconnection
or perhaps amplified by some mechanism in internal shocks.
The magnetic energy may have significantly dissipated
in Rabbit so that particles could diffuse out more efficiently, as speculated
based on the TeV emission outside the X-ray PWN.
There are other PWNe whose TeV emission is more extended than the radio/X-ray
emitting regions. It will be intriguing to see if these PWNe also exhibit a flat radial
profile of their X-ray photon index with deep X-ray observations.
Further multi-wavelength observations of other PWNe detected in the TeV band,
including NuSTAR hard X-ray observations, will be presented in our forthcoming papers
offering a good opportunity of exploring the PWN origin of Galactic PeVatrons  \citep{Mori2021}.   

\bigskip

\acknowledgments
This work used data from the NuSTAR mission, a project led by the California Institute of Technology,
managed by the Jet Propulsion Laboratory, and funded by NASA. We made use of the NuSTAR Data
Analysis Software (NuSTARDAS) jointly developed by the ASI Science Data Center (ASDC, Italy)
and the California Institute of Technology (USA).
This research was supported by Basic Science Research Program through
the National Research Foundation of Korea (NRF)
funded by the Ministry of Science, ICT \& Future Planning (NRF-2022R1F1A1063468).
Support for this work was partially provided by NASA through NuSTAR Cycle 6
Guest Observer Program grant NNH19ZDA001N.
 SSH acknowledges support from the Natural Sciences and Engineering Research Council
of Canada (NSERC) through the Discovery Grants and Canada Research Chairs programs and
from the Canadian Space Agency (CSA).
We thank the referee for a detailed reading of the manuscript and constructive comments
that helped strengthen the paper.
\vspace{5mm}
\facilities{CXO, XMM-Newton, NuSTAR, Fermi/LAT}
\software{HEAsoft (v6.29; HEASARC 2014), CIAO \citep[v4.13;][]{fmab+06},
XMM-SAS \citep[v20180620;][]{xmmsas17}, XSPEC \citep[v12.12;][]{a96},
FermiPy \citep[v1.0.1;][]{Wood2017}}

\bibliographystyle{apj}
\bibliography{ms.bbl}

\end{document}